\documentclass[12pt]{article}
\newcommand{\beq}{\begin{equation}}
\newcommand{\eeq}{\end{equation}}
\newcommand{\beqa}{\begin{eqnarray}}
\newcommand{\eeqa}{\end{eqnarray}}
\newcommand{\lam}{\lambda}
\newcommand{\ph}{\varphi}
\newcommand{\eps}{\varepsilon}
\newcommand{\bd}[1]{ \mbox{\boldmath $#1$}  }
\newcommand{\ap}{\approx}
\usepackage{epsfig}
\begin{document}
\def\ii{\'\i}

\title{
Collective Modes of Tri-Nuclear Molecules}

\author{P. O. Hess$^1$, \c S. Mi\c sicu$^2$, W.Greiner$^3$ and W.Scheid$^4$\\
{\small\it
$^1$ Instituto de Ciencias Nucleares, UNAM, Circuito Exterior, C.U.,} \\
{\small\it A.P. 70-543, 04510 M\'exico, D.F., Mexico} \\
{\small\it
$^2$ National Institute for Nuclear Physics, Bucharest, P.O.Box MG6,
Romania} \\
{\small\it
$^3$ Institut f\"ur Theoretische Physik, J.W.v.-Goethe Universit\"at,} \\
{\small\it Robert-Mayer-Str. 8-10, 60325 Frankfurt am Main, Germany} \\
{\small\it $^4$ Institut f\"ur Theoretische Physik, Julius Liebig Universit\"at} \\
{\small\it Heinrich-Buff-Ring 16, 35392 Giessen, Germany}
}
\maketitle 

\abstract{
A geometrical model for tri-nuclear molecules is presented. An
analytical solution is obtained provided the nuclei, which are taken to be 
prolately deformed, are connected in line to each other. Furthermore, the 
tri-nuclear molecule is composed of two heavy and one light cluster, the later 
sandwiched between the two heavy clusters.
A basis is constructed in which Hamiltonians of more general configurations
can be diagonalized. 
In the calculation of the interaction between the
clusters higher multipole deformations are taken into account, including the 
hexadecupole one. A repulsive nuclear core is introduced in the potential in order 
to insure a quasi-stable configuration of the system. The model is applied
to three
nuclear molecules, namely $^{96}$Sr $+$ $^{10}$Be $+$ $^{146}$Ba,
$^{108}$Mo $+$ $^{10}$Be $+$ $^{134}$Te and 
$^{112}$Ru $+$ $^{10}$Be $+$ $^{130}$Sn.}

\section*{1. Introduction}
In a recent experiment \cite{hamilton} the cold fission decay of
$^{252}$Cf into three clusters was investigated. In coincidence
measurements the three participants were identified as being
$^{96}$Sr, $^{10}$Be and $^{146}$Ba. In reanalyzing the data, two
further systems where discovered \cite{hamilton2}, namely
$^{90}$Y $+$ $^{142}$Cs and $^{108}$Mo $+$ $^{134}$Te, all with $^{10}$Be
as the third particle. 
Surprisingly
the data suggest that
the transition from the first excited $2^+$ state to the ground
state in $^{10}$Be was not Doppler broadened as one would expect if the
system immediately separates into three clusters and the Be nucleus deexcites in 
flight. In addition one observes that the transition energy of 3368 keV in the 
$^{10}$Be nucleus interacting with $^{96}$Sr and $^{146}$Ba is
probably
by about 6 keV
smaller than for the free $^{10}$Be nucleus. The transition energy decreases
further for the other two systems, being largest when both heavy
clusters are spherical. 
The heuristic explanation is that the average shell model 
frequency in presence of the two heavy clusters is modified. The 
influence of both clusters leads to a softening of the $^{10}$Be
potential and thus to a somewhat smaller transition energy.
The largest overlap, 
i.e. the strongest change in the average shell model frequency, of one heavy 
cluster with $^{10}$Be is obtained for a spherical deformation. 
Though, one has to wait for further experimental confirmation and better
statistics, the indication of the possible existence of three-cluster
molecules calls for a theoretical description and prediction of their
structure.

The interpretation of the above observation is the
probable
existence of a nuclear molecule
\cite{scheid} with a half life larger than $10^{-13}$sec \cite{hamilton}.
Such large lifetimes would open up the possibility of a spectroscopy of giant
nuclear molecules.
In \cite{letter} we proposed a phenomenological, geometrical model for
the system of three clusters, two heavy ones and a light in the middle.
The model was restricted to butterfly and belly-dancer modes only and
the $^{10}$Be cluster was assumed to be spherical.
However, in general the light
cluster can be deformed and its effect must be studied. Also the inclusion
of $\beta$ and $\gamma$ vibrations must be considered. Because the
light cluster in \cite{letter} was considered to be spherical, the
stiffness of the butterfly motion 
is mainly determined 
by the monopole part of the Coulomb repulsion between the heavy fragments 
\cite{letterdouble}. 
In the case when the light fragment is deformed, this is no longer true and one 
has to determine explicitely the change of the nuclear and Coulomb interaction 
between the light and heavy clusters as a function of the inclination angle.
The effective potential is determined via a double folding calculation
and is described in Ref. \cite{double}. In Ref. \cite{letterdouble}
the nuclear potential was taken into account, including also 
multipoles higher than the monopole and quadrupole ones.

In this contribution we present the details of the model proposed in
Ref. \cite{letter} which itself is an extension 
of Ref. \cite{belly} presented for two clusters. The molecule exhibits butterfly 
and belly-dancer modes, $\beta$- and $\gamma$-vibrations of the two
clusters. This picture can be extended straight\-forwardly
to three clusters using the formulas as given in Ref. \cite{belly}.
It is especially easy for the case when the
two big clusters (for example, the $^{96}$Sr and $^{146}$Ba) are connected via
a smaller spherical nucleus ($^{10}$Be). The situation is illustrated
in Fig. 1, where the main dynamical variables, with a spherical
cluster in the middle, are indicated. In this paper we will extend the model
to include also $\beta$ and $\gamma$ vibrations, with the addition
of a deformed light cluster.

In Fig.2 possible vibrational modes are indicated, when the light cluster is 
deformed. The first case corresponds to the above mentioned butterfly mode, whereas
the second one is the antibutterfly mode and is not investigated in this 
contribution. We will see that for the butterfly motion the dominant contribution 
comes from the movement of the center of masses and, contrary to the two cluster 
case, deformations play a minor role, except in fixing the length of the axis in 
Fig. 1 and 2. Also the methods presented in this paper can be readily extended to 
this antibutterfly mode. 

In order to keep the problem tractable, the main assumptions are that the light 
cluster is sandwiched in-between the heavy nuclei and the inclination angles of 
the clusters are small with respect to the molecular (fission) $z$-axis, connecting 
both heavy clusters. In such a linear chain configuration the total potential of 
the lighter cluster has an absolute minimum on the axis joining the three
fragments. As can be seen in Fig.3, for a given distance $d$ between the 
tips of the two heavier fragments, there is a point on the fission axis
$z$, where the forces exerted by the heavy fragment 1 on the light
cluster are canceled by the forces exerted by the heavy fragment 2.
This is the so-called electro-nuclear saddle point \cite{misicim99}.
The details of the potential calculation are explained in section 3. For
the moment we just note that the result of Fig.3 was obtained using 
a strong nuclear repulsive force, between the fragments in order to avoid
their mutual overlap. 

The paper is organized as follows:  In section 2 the general formulation
of a three cluster molecule is presented. For practical purposes we then
restrict to
the case where the three clusters are in touch and all three can be
deformed. This section will be divided into one where no $\beta$ and
$\gamma$ vibration is included and a second one where these degrees
of freedom are considered, too.
In section 3 the derivation of the nuclear and Coulomb potentials is given.
In section 4 the model is applied to the three systems:
$^{96}$Sr + $^{10}$Be + $^{146}$Ba,
$^{108}$Mo $+$ $^{10}$Be $+$ $^{134}$Te and 
$^{112}$Ru $+$ $^{10}$Be $+$ $^{130}$Sn.
All these splittings are a product of the ternary cold fission of
$^{252}$Cf. The structure of the participants is discussed. Finally, in
section 5 conclusions are drawn.  

\section*{2. The general formalism}
The motion of the three clusters can be divided into the rotations of the
individual clusters plus the motion of their center of masses with respect
to each other. The part of the Hamiltonian which describes the individual rotations
can be read off from Eqs. (49) and (50) of Ref. \cite{belly}. The part which 
changes originates from the motion of the center of masses. Therefore, the 
discussion concentrates first on the motion of the three center of masses and is 
independent of considering either of the two modes described in Fig. 2.

In order to separate the center-of-mass motion, the following coordinates are 
introduced 

\begin{eqnarray}
\bd{r} & = & \bd{r_2}-\bd{r_1} \nonumber \\
\bd{\xi} & = & \frac{m_1\bd{r_1}+m_2\bd{r_2}}{m_1+m_2}-\bd{r_3}
\label{coord}
\end{eqnarray}
where $m_k = A_k m$, $A_k$ is the number of nucleons of cluster no. $k$ and
$m$ the nucleon mass. The first coordinate ($\bd{r}$) describes the relative 
distance of the two heavy fragments while the second one ($\bd{\xi}$) the
distance of the third, lightest cluster to the center of mass of the first
two. The kinetic energy, excluding the motion of the total center of mass,
acquires the form
\beq
T = {1\over 2}\mu_{12}{\dot{\bd{r}}}^2
+ {1\over 2}\mu_{(12)3}{\dot{\bd{\xi}}}^2
+ {1\over 2}{^t}\bd{\omega}_1\bd{\cal J}_1\bd{\omega}_1
+ {1\over 2}{^t}\bd{\omega}_2\bd{\cal J}_2\bd{\omega}_2
+ {1\over 2}{^t}\bd{\omega}_3\bd{\cal J}_3\bd{\omega}_3     
\label{hamcm}
\eeq   
where $ \mu_{12}=\frac{m_1m_2}{2(m_1+m_2)}$ and 
$\mu_{(12)3}=\frac{m_3(m_1+m_2)}{2(m_1+m_2+m_3)}$.
The first term in (\ref{hamcm}) describes the kinetic energy of the
two heavy clusters with respect to each other and the second term
the kinetic energy of the third cluster with respect to a mass $(m_1+m_2)$
at the center of mass of the first two clusters. The mass factors describe
the reduced mass for each case. The term proportional to $\dot{\bd{r}}^2$
has the same form as for the two cluster case and thus is already
included in the considerations of Ref. \cite{belly}. The last three terms in 
eq.(\ref{hamcm}) are describing the rotational motion of the three clusters with
angular velocities $\bd{\omega}_{1,2,3}$, referred to the laboratory frame. The 
inertia tensors $\bd{\cal J}_i$ are defined in the intrinsic frame such that
in the absence of $\beta$ and $\gamma$ vibrations
the only non-vanishing components are the first two diagonal terms, 
$(\bd{\cal J}_i)_{11}=(\bd{\cal J}_i)_{22}\equiv J_i$, the quantum rotation around 
the symmetry axis of any of the two heavier fragments being discarded.
When $\beta$ and $\gamma$ vibrations are included there will be a contribution
to $(\bd{\cal J}_i)_{33}$ given by a $\gamma$ dependence \cite{greiner}. 

The second term in Eq. (\ref{hamcm}) needs more attention. 
In Fig. 4 the three center of masses are plotted and the relevant coordinates
are indicated. The projection of the vector $\bd{\xi}$ onto the relative distance 
vector, denoted by $\xi_z$, and its perpendicular component along the $x$-axis
($\xi_x$)  are given by

\begin{eqnarray}
\xi_z & = & \frac{A_2}{(A_1+A_2)}r -
r_{13}\cos\eps \nonumber \\
\xi_x & = & r_{13} \sin\eps
\label{xizx}
\end{eqnarray}
where $r_{13}$ is the distance between the cluster no. 1 and 3 and
$\eps$
is the angle between the axis connecting cluster 1 and 3 to the vector
$\bd{r}$. Note, that this angle is not necessarily the same as the
inclination angle of the intrinsic $z$-axis of cluster no. 1 to the vector
$\bd{r}$. Because the molecular plane is defined by the three center of
masses of the clusters the spherical components $\xi_{\pm 1}$ are
the same up to a sign (see Appendix A and note the change in sign
in the definition of the spherical components, due to convenience,
compared to the usual definition \cite{edmonds}). The $\bd{\xi}$ contribution of 
the kinetic energy is obtained by rotating into the molecular system and then
substituting the expressions for the spherical components of the
vector $\bd{\xi}$. In order to keep the procedure manageable, we assumed
that the angle $\eps$ is small, i.e. the light, third cluster is
not far away from the axis connecting the two heavy clusters.
Furthermore, we assume that the first cluster, supposed to be the lightest
one of the two heavy clusters, touches the third one and the third
cluster touches the second one. The situation is illustrated in
Fig. 5 with a certain exaggeration as concerned to the distance of the
third cluster to the axis ($\bd{r}$) connecting the heavy clusters.
These assumptions exclude the anti butterfly mode for which the relation of the 
angles change. The procedure for that mode, however, is completely analogous.

In Fig. 5 the relation of the angle $\varphi_3$ to $\varphi_1$ and $\varphi_2$ can 
be read off with the supposition that the three clusters are connected. 
Assuming small angles, the result is
\begin{equation}
\varphi_3 \approx \sin\varphi_3 = 
\frac{1}{2R_3} (R_2\sin\varphi_2-R_1\sin\varphi_1)
\approx \frac{1}{2R_3} (R_2\varphi_2-R_1\varphi_1)
\label{phi3}
\end{equation}
In the above formula we suppose that the heavy cluster 2 has a larger shift on
the $x$-direction than the lighter one. In principle the two angles $\varphi_1$ 
and $\varphi_2$ could be treated now as independent with $\varphi_3$ constrained 
by Eq. (\ref{phi3}). However, the problem would get too complicated, implying 
coupling terms between the $\varphi_1$ and $\varphi_2$ motion. Supposing that the 
cluster in the middle is small, the relation between $\varphi_1$ and $\varphi_2$ 
should not differ much from the case when the small cluster is spherical 
\cite{letter}. We use, therefore, the same relation 
$\varphi_2$ $\approx$ $\frac{(R_1+R_3)}{(R_2+R_3)}\varphi_1$ as for a spherical 
cluster and substitute it into Eq. (\ref{phi3}) resulting in 
($\varphi_1\approx \eps$)
\begin{equation}
\varphi_3 \ap {1\over 2}\frac{R_2-R_1}{R_2+R_3}\eps ~~~.
\label{phi3-f}
\end{equation}
One possibility to relax this constraint is to expand the general motion around 
the point where condition (\ref{phi3-f}) holds and diagonalizes the Hamiltonian in 
this basis. This standpoint will be adopted in the present paper.

For the case of connected clusters the distance between center of masses of the 
third cluster and nucleus no.1, $R_{13}$, can be approximated by $(R_1+R_3)$ in 
the linear, unperturbed configuration. In the perturbed case, when the molecule 
ceases to be linear, $r_{13}$ is modified by 
the amount
\beq
\delta r_{13} \ap -{1\over 8}\frac{R_1R_3}{R_1+R_3}
\left (\frac{R_1+R_2+2R_3}{R_2+R_3}\right )^2\eps^2
\label{r13}
\eeq
provided $\eps$ is small,
Similarly we get the variation of $r_{23}$ 
\beq
\delta r_{23} \ap -{1\over 8}\frac{R_2R_3}{R_2+R_3}
\left (\frac{R_1+R_2+2R_3}{R_2+R_3}\right )^2\eps^2
\label{r23}
\eeq

At equilibrium ($\eps$=0), the distance $r$, between the center of masses of the
two heavy clusters is given by $R_0\equiv R_1+R_2+2R_3$. When $\eps\neq 0$, and 
taking into account terms up to second order in $\eps$, $r$ changes to 

\begin{equation}
r \ap (R_1+2R_3+R_2) \left( 1-{1\over 2}\frac{R_1+R_3}{R_2+R_3}\eps^2 \right) 
+\delta r_{13} + \delta r_{23}.
\label{r12}
\end{equation}

Therefore, an increase in $\eps$,
changes $r$, $r_{13}$ and $r_{23}$ with a correction in the moment of inertia of
the order of $\eps^2$ and higher.
At their turn, the components $\xi_z$ and $\xi_x$ acquire the form
\begin{eqnarray}
\xi_z & \approx &
\left[ \frac{A_2}{(A_1+A_2)}(R_1+2R_3+R_2)
- (R_1+R_3) \right]  \nonumber \\
& + & \frac{A_2}{A_1+A_2}\delta r_{12}-
           \delta r_{13}+\frac{(R_1+R_3)}{2}\eps^2\nonumber \\
  & = & \xi_0 + \delta\xi_z\nonumber\\
\xi_x & \approx & (R_1+R_3) \eps + \delta\xi_x \hspace*{0.5cm} .
\label{xizx2}
\end{eqnarray}
where $\delta\xi_z$ and $\delta\xi_x$ give the contributions due to
changes in the relative distance of the clusters 1 to 3 and 2 to 3, i.e. they 
describe a mode of the stretching vibrations. In the unperturbed  configuration
\begin{equation}
\xi_0 =\frac{A_2(R_2+R_3)-A_1(R_1+R_3)}{A_1+A_2} 
\end{equation}
Again, for simplicity we assume that the relative vibrations are along the 
molecular $z$ axis, i.e. $\delta\xi_x=0$. If this assumption is not made, there 
will be contributions of the type $\delta\dot\xi_x\delta\dot\xi_z$ in the kinetic 
energy. One has then to construct the basis first, where this coupling term is 
absent, and afterwards to diagonalize the complete Hamiltonian in this basis. 
Note that the model is quite crude and other coupling terms, like rotation-
vibration interactions, have been neglected. Consequently, the spectrum obtained 
in the model will be a first approximation and the assumptions made above are 
justified only  in this context.

The radial mode $\xi$ along the $z$ axis describes the motion of
the two heavy clusters in a common direction and of the light cluster
in the opposite direction (see Ref. \cite{letter}).
The other radial mode comes from changes
in $r$ and describes a vibration of the two heavy clusters with respect to
each other while the small cluster does not move. These types of
vibrations will be included in the complete Hamiltonian.
Inserting (\ref{xizx2})
into the kinetic energy for the $\bd{\xi}$ motion, using $\delta\xi_x=0$,
we obtain(see Appendix A)

\begin{eqnarray}
T_{\xi} & = & {1\over 2}\mu_{(12)3}\left\{(R_1+R_3)^2 {\dot\eps}^2 + 
\xi_0^2(\omega_1^{\prime ~2} + \omega_2^{\prime ~2}) 
+(R_1+R_3)^2\eps^2\omega_3^{\prime ~2}\right.\nonumber\\ 
&+& \left.2 \xi_0 (R_1+R_3)(\dot\eps \omega_2^{\prime} -
\eps\omega_1^{\prime}\omega_2^{\prime})\right\} ~~~~~~~,
\label{tkinxi}
\end{eqnarray}
with $\omega^{\prime}_k$ being the angular velocity around the $k$'th
molecular axis ($1=x$, $2=y$ and $3=z$) and $\xi$ is a shorthand notation
for $\delta\xi_z$. From here on, when we refer to the total contribution,
including the rotational one, we continue to denote it by $\bd{\xi}$ and
the pure motion along the $z$ axis is denoted by $\xi$.

The kinetic energy of the $r$ and $\xi$ stretching vibrations
along the molecular $z$ axis is described by

\begin{eqnarray}
T_r & = & {1\over 2}\mu_{12} \dot{r}^2 + {1\over 2}\mu_{(12)3}\dot\xi^2 ~~~,
\label{trel}
\end{eqnarray}
where the last term comes from Eq. (\ref{tkinxi}).
The coordinates $r$ and $\xi$ are related to the relative distances
$r_{13}= (z_3-z_1)$ and $r_{23}=(z_2-z_3)$ via

\begin{eqnarray}
(z_2-z_3) & = & \frac{m_1}{m_1+m_2}r +\xi \nonumber \\
(z_3-z_1) & = & \frac{m_2}{m_1+m_2}r -\xi ~~~.
\end{eqnarray}
The corresponding potential is given by

\begin{eqnarray}
V_r & = &
\frac{C_{13}}{2} ((r_3-r_1)-r_{13,0})^2 +
\frac{C_{23}}{2} ((r_2-r_3)-r_{23,0})^2
\end{eqnarray}
where $r_{13,0}$ and $r_{23,0}$ are the equilibrium positions
$r_{13,0}=(R_1+R_3)$ and $r_{23,0}=(R_2+R_3)$ respectively. Note that
in our picture cluster no. 1 is to the left and no. 2 to the right and
that the coordinates refer to the distance along the intermolecular axis.

Up to know we considered the $\bd{r}$ and the $\bd{\xi}$ motion only.
The contributions coming from the deformation of the individual clusters
can be read off from Ref. \cite{belly} and from there the general
kinetic energy can be constructed.

In the next subsection we will consider two different cases mentioned
in the introduction. In the first one the contributions of $\beta$ and $\gamma$ 
vibrations are excluded and in the second one they are included. 
In both cases the static deformation in the ground state
is assumed to be prolate. This restriction is governed by the
necessity to keep the problem solvable otherwise the complicated form
of the Hamiltonian would prevent an analytical solution. In case
a triaxial nucleus is present the procedure outlined is strictly
speaking not valid, but an approximation of the triaxial nucleus by an
axial symmetric 
one
would do the job.

\subsection*{2.1 Without $\beta$ and $\gamma$ vibrations}
With the assumption made in the preceeding section
the total kinetic energy is given by

\begin{eqnarray}
T & = & \frac{1}{2}\Theta_{11} (\omega_1^{\prime ~ 2} +
\omega_2^{\prime ~ 2}) + {1\over 2}\Theta_{33} \omega_3^{\prime ~ 2}
-\Theta_{13}\eps\omega_1^{\prime}\omega_3^{\prime}+ 
{1\over 2}\Theta_\eps {\dot\eps}^2 + \Theta_{2\eps} \dot\eps\dot\omega_2^{\prime}
\nonumber \\
& + &{1\over 2}\mu_{12}\dot{r}^2 + {1\over 2}\mu_{(12)3}\dot\xi^2
\label{ttotsin}
\end{eqnarray}
with

\begin{eqnarray}
\Theta_{11} & \approx & J_1 + J_2 + J_3+ \mu_{(12)3}\xi_0^2+ \mu_{12}R_0^2 
\nonumber \\
\Theta_{33} &  \approx & \left ( J_1 + J_2\frac{(R_1+R_3)^2}{(R_2+R_3)^2} +
J_3\frac{(R_1-R_2)^2}{4(R_2+R_3)^2} + \mu_{(12)3}(R_1+R_3)^2\right )\eps^2 
\nonumber \\
\Theta_{13} & \approx & J_1+ J_2\frac{(R_1+R_3)}{(R_2+R_3)} +
J_3\frac{(R_1-R_2)}{2(R_2+R_3)} + \mu_{(12)3} \xi_0(R_1+R_3)
\nonumber \\
\Theta_{\eps} & = & {1\over \eps^2}\Theta_{33}
\nonumber \\
\Theta_{2\eps} & = &\Theta_{13}
\label{theta1}
\end{eqnarray}
For nearly symmetric heavy clusters, not too deformed and a small value of 
$\frac{m_3}{m_1+m_2}$ the coupling is small compared to the diagonal terms of the 
moment of inertia and thus can be neglected, as for the other contributions 
(see Ref. \cite{belly}). Also the deformation dependent part in $\Theta_{11}$ can 
be neglected with respect to the last term, as was done also in Ref. \cite{belly}.
However, for very asymmetric heavy clusters we cannot neglect any more the 
contribution of $\Theta_{2\eps}$, except for small $\frac{m_3}{m_1+m_2}$. 
In this case, one has to diagonalize this term in the basis with  
$\Theta_{2\eps}=0$, which is in the same spirit as for the coupling terms in the 
radial vibrations. The basis will be discussed further below.

For the potential we assume a quadratic behaviour in $\eps$, $r$ and $\xi$, i.e.

\begin{eqnarray}
V & = & \frac{C_{\eps}}{2} \eps^2
+ \frac{C_r}{2} \bar{r}^2 +\frac{C_\xi}{2}(\xi -\xi_0)^2 \hspace*{0.5cm} ,
\label{potsin}
\end{eqnarray}
where $\bar{r}=(r-r_0)$ with $r_0$ being the equilibrium position of
the nuclear molecule.
The parameters $C_r$ and $C_\xi$ are related to $C_{13}$ and $C_{23}$ via
\begin{eqnarray}
C_r & = & \frac{(m_2^2C_{13}+m_1^2C_{23})}{(m_1+m_2)^2} \nonumber \\
C_\xi & = & C_{13}+C_{23} ~~~.
\label{cr&cxi}
\end{eqnarray}
Other crossing terms of the type $r\xi$, $r$ and $\xi$ also appear, which vanish 
for a symmetric dinuclear sub-system, formed from clusters 1 and 2. We assume that 
in general the microscopic interaction is such that also for non-symmetric 
clusters the coupling terms can be canceled or made small, which is obviously
a simplification.

In what follows, we quantize the Hamiltonian with the kinetic energy
given in (\ref{ttotsin}) taking into account the contribution of the
coupling of the $\eps$ and $\omega_2$ motion, whose origin is the
Coriolis force. The quantization procedure is explained in Ref.
\cite{greiner} and was also used in Ref. \cite{belly}.

Then a metric can be identified via

\begin{equation}
2Tdt^2 = \sum_{\mu\nu}g_{\mu\nu} dq_\mu dq_\nu
\label{ds2}
\end{equation}
where $T$ is the total kinetic energy and
$dq_\mu$ is a short hand notation for the variables appearing in
our model. The quantized kinetic energy is given by

\begin{eqnarray}
\hat{T} & = & -\frac{\hbar^2}{2}\sum_{\mu\nu} \frac{1}{\sqrt{g}}
\frac{\partial}{\partial q_\mu} \sqrt{g}(g^{-1})_{\mu\nu}
\frac{\partial}{\partial q_\nu}
\label{tquant}
\end{eqnarray}
with $g=det(g_{\mu\nu})$. The volume element is

\begin{eqnarray}
d\tau & = & \sqrt{g}dq_1~ ... ~ dq_n = \sqrt{g}d\tau^{\prime}
\hspace*{0.5cm} .
\label{volume}
\end{eqnarray}
It is more convenient to use as volume element $d\tau^{\prime}$ instead.
We have to redefine the Hamiltonian $\hat{H}$ and
the wave function $\psi$ of the original Schr\"odinger equation via

\begin{eqnarray}
\hat{H} & \rightarrow & \hat{H}^{\prime} = g^{\frac{1}{2}}\hat{H}
g^{-\frac{1}{2}} \nonumber \\
\psi &\rightarrow & \phi = g^{\frac{1}{2}} \psi
\label{newham}
\end{eqnarray}
so that

\begin{eqnarray}
\int \psi^* \hat{H} \psi d\tau = \int \phi^* \hat{H}^{\prime}
\phi d\tau^{\prime} \nonumber
\end{eqnarray}
holds. The potential commutes with $g$ because it does not contain
differential operators. The kinetic energy simplifies to

\begin{eqnarray}
\hat{T} & = & -\frac{\hbar^2}{2} \sum_{\mu\nu} \frac{\partial}{\partial q_\mu}
(g^{-1})_{\mu\nu} \frac{\partial}{\partial q_\nu} + V_{add}
\nonumber \\
V_{add} & = & -\frac{\hbar^2}{8}\sum_{\mu\nu} \left\{
\frac{3}{4}(g^{-1})_{\mu\nu}\frac{1}{g^2}\frac{\partial g}{\partial q_\nu}
\frac{\partial g}{\partial q_\nu} - \frac{1}{g}
\frac{\partial (g^{-1})_{\mu\nu}}{\partial q_\mu}
\frac{\partial g}{\partial q_\nu} - \right. \nonumber \\
& & (g^{-1}_{\mu\nu}) \frac{1}{g}
\left. \frac{\partial^2g}{\partial q_\nu \partial q_\mu} \right\}
\hspace*{0.5cm} .
\label{newt}
\end{eqnarray}

The Hamiltonian composed by the kinetic energy (\ref{ttotsin}) and the
potential (\ref{potsin}) is quantized in this manner, resulting for
the kinetic energy in

\begin{eqnarray}
T &  \approx & 
\frac{\hbar^2({\hat I}^2 - {\hat I}^{\prime 2}_3)}
{2(\Theta_{11}-\frac{\Theta_{13}^2}{\Theta_{\eps}})} +
\frac{\hbar^2 {\hat I}_3^{\prime 2}}
{2(\Theta_{\eps}-\frac{\Theta_{13}^2}{\Theta_{11}})\eps^2} 
 -\frac{\hbar^2}{2(\Theta_{\eps}-\frac{\Theta_{13}^2}{\Theta_{11}})}
\left (\frac{\partial^2}{\partial\eps^2} + {1\over 4\eps^2}\right )
\nonumber\\
&-& \frac{1}{2}\mu_{12} \frac{\partial^2}{\partial  \bar{r}^2}
- \frac{1}{2}\mu_{(12)3}\frac{\partial^2}{\partial  {\xi}^2}
\nonumber \\
& + & \frac{1}{(\frac{\Theta_{11}\Theta_{\eps}}{\Theta_{13}}-\Theta_{13})}
\left\{ \frac{\hbar^2}{\eps}{\hat I}_1^{\prime}{\hat I}_{3}^{\prime}
-\hbar\hat{I}_2^\prime \left (\frac{\hbar}{i}\frac{\partial}{\partial\eps}\right )
\right\}
\label{hamquant1}
\end{eqnarray}
In the following discussion we will skip the term in the parenthesis
$\{ ... \}$ which has to be treated as a 
perturbative interaction
and the corrections given by 
$\frac{\Theta_{13}^2}{\Theta_{\varepsilon}\Theta_{11}}$, which is
small compared to one for the molecular systems
studied in this paper. In very assymmetric systems, however, both terms
have to be included and diagonalized in the basis which will be constructed
in what follows.
The terms which do not contain a derivative come from the additional
potential of Eq. (\ref{newt})

Neglecting the term with the parenthesis $\{ ... \}$ in Eq. (\ref{hamquant1})
and corrections of the order of $\frac{\Theta_{13}^2}{\Theta_{11}}$ and
$\frac{\Theta_{13}^2}{\Theta_{\varepsilon}}$,
the corresponding static Schr\"odinger equation can be solved with
the ansatz

\begin{equation}
\phi = D^{I^*}_{MK} (\vartheta) \chi_{K, n_\eps} (\eps) g_{n_r} (\-r)
\label{ansatz1}
\end{equation}
were $D^{I^*}_{MK} (\vartheta)$ is the Wigner $D$-matrix \cite{edmonds},
$g_{n_r} (\-r)$ is the one dimensional harmonic oscillator for the
relative motion and $\chi_{K,n_\eps}$ is the solution of the
differential equation

\begin{eqnarray}
& \left[ -\frac{\hbar^2}{2\Theta_{\eps}}\frac{\partial}
{\partial \eps^2} + {1\over \Theta_{\eps}}(K^2 -\frac{1}{4})
\frac{\hbar^2}{2\eps^2} +\frac{C_{\eps}}{2} \eps^2 \right.  & \nonumber \\
& \left. + \left( \frac{\hbar^2}{2\Theta_{11}}[I(I+1)-K^2] 
+ \hbar\omega_r(n_r+\frac{1}{2})+ \hbar\omega_\xi (n_\xi +\frac{1}{2})
-E\right) \right]\phi & =  0 \hspace*{0.5cm} ,
\label{chi}
\end{eqnarray}
where $I$ is the total spin, $K$ its projection to the molecular $z$ axis,
$\hbar\omega_r$ $=$ $\hbar\sqrt{\frac{C_r}{\mu_{(12)}}}$,
$\hbar\omega_\xi$ $=$ $\hbar
\sqrt{\frac{C_\xi}{\mu_{3(12)}}}$ and
$E$ the total energy of the state $\phi$.

This equation can be solved with the solution given by (see also Ref.
\cite{greiner})

\beq
\chi_{K, n\eps}(\eps) = 
\frac{\left\{ \lambda^{l_k+\frac{3}{2}}\Gamma (l_K+\frac{3}{2}+n_{\eps})
\right\}^{\frac{1}{2}}}{(n_{\eps}!)^{1\over 2}\Gamma (l_K+\frac{3}{2})}
\eps^{l_K+1} e^{-\frac{1}{2}\lambda\eps^2} ~ 
_1F_1(-n_{\eps}, l_K+\frac{3}{2};\lambda\eps^2)
\hspace*{0.5cm} ,
\label{sol1}
\eeq
where $\lambda^2$ $=$ $\frac{\Theta_{\eps}C_{\eps}}{\hbar^2}$,
$l_K = \mid K \mid - {1\over 2}$
and $_1F_1(...)$ is the confluent hypergeometric function.

The total energy is given by

\begin{equation}
E = \frac{\hbar^2}{2\Theta_{11}} \left[ I(I+1) - K^2 \right]
+\hbar\omega_r (n_r + \frac{1}{2}) + \hbar\omega_{\eps}
(\mid K \mid +2n_{\eps}+\frac{3}{2}) 
\label{energy1}
\end{equation}
and $\hbar\omega_{\eps}$ $=$
$\hbar\sqrt{\frac{C_{\eps}}{\Theta_{\eps}}}$.

When there are two identical clusters the wavefunction has still to be
symmetrized with respect the interchange of cluster no. 1 and 2. For details
on how the variables change see Ref. \cite{belly} at least for the two heavy
clusters. For the light cluster the variables transform in the same way as
indicated by the second cluster except for the change of indices
from $2$ to $3$.
With respect to the variable $\eps$ the transformation is 
$\eps \rightarrow -\eps$.

\subsection*{2.2 Including $\beta$ and $\gamma$ vibrations}
The extension to $\beta$ and $\gamma$ vibrations is straightforward.
Including also the rotation around the intrinsic $z$ axis of
axial symmetric nuclei, given by
$\Phi_k$ ($k=1,2,3$), leads to a very complicated form.
Additionally we have to assume that the clusters are prolately
deformed. Otherwise a complex coupling between the rotations
around the $x$, $y$ and $z$ will appear. The $\beta$ and $\gamma$
vibrational variables of the $k$'th cluster are defined by

\begin{eqnarray}
\bar{\beta}_k & = & a_0^k-\beta_{0k} \nonumber \\
\eta_k & = & a_2^k ~~~,
\end{eqnarray}
where $a_0^k$ and $a_2^k$ are the components of the quadrupole
deformation variable of the $k$'th cluster with respect to the
principal axes.
Using Eq. (50) of the second paper in
Ref. \cite{belly} for the moments of inertia plus the
corrections discussed further up, the Hamiltonian acquires the form

\begin{eqnarray}
T & = & \frac{\hbar^2({\hat I}^2 - {\hat I}^{\prime 2}_3)}
{2(\Theta_{11}-\frac{\Theta_{13}^2}{\Theta_{\eps}})} +
\frac{\hbar^2 {\hat I}_3^{\prime 2}}
{2(\Theta_{\eps}-\frac{\Theta_{13}^2}{\Theta_{11}})\eps^2} 
 -\frac{\hbar^2}{2(\Theta_{\eps}-\frac{\Theta_{13}^2}{\Theta_{11}})}
\left (\frac{\partial^2}{\partial\eps^2} - {1\over 4\eps^2}\right )\nonumber\\
& - & \frac{\hbar I^\prime_3}{\Theta_{\eps}\eps^2} (\frac{\hbar}{i}
\frac{\partial}{\partial\Phi_1} + \frac{\hbar}{i}
\frac{\partial}{\partial\Phi_2} + \frac{\hbar}{i}
\frac{\partial}{\partial\Phi_3}) \nonumber \\
& & -\frac{\hbar^2}{2\Theta_{\eps}\eps^2} \left(
\frac{\partial^2}{\partial\Phi_1^2} +
\frac{\partial^2}{\partial\Phi_2^2} +
\frac{\partial^2}{\partial\Phi_3^2} \right) \nonumber \\
& & -\frac{\hbar^2}{\Theta_{\eps}\eps^2} \left(
\frac{\partial^2}{\partial\Phi_1\partial\Phi_2} +
\frac{\partial^2}{\partial\Phi_1\partial\Phi_3} +
\frac{\partial^2}{\partial\Phi_2\partial\Phi_3} \right) \nonumber \\
& & -\frac{\hbar^2}{2B_1}\frac{\partial^2}{\partial^2\bar{\beta}_1^2} 
-\frac{\hbar^2}{2B_2}\frac{\partial^2}{\partial^2\bar{\beta}_2^2} 
-\frac{\hbar^2}{2B_3}\frac{\partial^2}{\partial^2\bar{\beta}_3^2} 
\nonumber \\
& & -\frac{\hbar^2}{16B_1\eta_1^2} -\frac{\hbar^2}{16B_3\eta_3^2}
-\frac{\hbar^2}{16B_3\eta_3^2} \nonumber \\
& & -\frac{\hbar^2}{4B_1}\frac{\partial^2}{\partial^2\eta_1^2} 
-\frac{\hbar^2}{4B_2}\frac{\partial^2}{\partial^2\eta_2^2} 
-\frac{\hbar^2}{4B_3}\frac{\partial^2}{\partial^2\eta_3^2} 
\nonumber \\
& & -\frac{\hbar^2}{16B_1\eta_1^2}\frac{\partial^2}{\partial\Phi_1^2} 
-\frac{\hbar^2}{16B_2\eta_2^2}\frac{\partial^2}{\partial\Phi_2^2} 
-\frac{\hbar^2}{16B_3\eta_3^2}\frac{\partial^2}{\partial\Phi_3^2}
\nonumber \\
& & -\frac{\hbar^2}{2\mu_{(12)}}\frac{\partial^2}{\partial\bar{r}^2}
-\frac{\hbar^2}{2\mu_{(12)3}}\frac{\partial^2}{\partial\xi^2}
\label{hamquant2}
\end{eqnarray}
where the variables $\bar{\beta}_k$ and $\eta_k$ ($k=1,2,3$)
describe the $\beta$ and $\gamma$ degree of freedom as defined
in Ref. \cite{greiner}.

The complete potential is given by

\begin{eqnarray}
V & = & \frac{C_{\eps}}{2} \eps^2
+ \frac{C_r}{2} \bar{r}^2 +\frac{C_\xi}{2}(\xi -\xi_0)^2 \nonumber \\
& & + C_{\eta_1}\eta_1^2 + C_{\eta_2}\eta_2^2 + C_{\eta_3}\eta_3^2
\nonumber \\
& & + \frac{C_{\beta_1}}{2}\bar{\beta}_1^2+ \frac{C_{\beta_2}}{2}\bar{\beta}_2^2 +
\frac{C_{\beta_3}}{2}\bar{\beta}_3^2
\end{eqnarray}
and the determinant $g$ is 

\begin{eqnarray}
g & = & 8^3 2^3 \Theta_{11}^2\Theta_{\eps}^2 B^3_1B^3_2B^3_3 
\eta^2_1\eta^2_2\eta^2_3 \eps^2 ~~~.
\end{eqnarray}
where $\Theta_{22}$ $\approx$ $\Theta_{11}$ was used. The factor
$2^3B_1^2B_2^2B_3^2$ $=$ $(2B_1)(2B_2)(2B_3)$ $B_1B_2B_3$ comes from the
$\beta$ and $\gamma$ part of the metric tensor $g_{\mu\nu}$.

Neglecting, as in the case without $\beta$ and $\gamma$ vibrations,corrections of 
the order of  $\frac{\Theta_{13}^2}{\Theta_{\varepsilon}\Theta_{11}}$
the static Schr\"odinger equation can be solved with the ansatz

\begin{eqnarray}
& &\phi (\vartheta_1, \vartheta_2, \vartheta_3, \eps , \bar{r}, \bar{\beta}_1,
\bar{\beta}_2,\bar{\beta}_3,\eta_1,\eta_2,\eta_3) =  
e^{i(K_1\Phi_1+K_2\Phi_2+K_3\Phi_3)} 
D^{I^*}_{MK} (\vartheta)\chi_{\tilde{K}, n\eps} (\eps) g_{n_r} (\bar{r})
\nonumber \\
& & g_{n_\xi}(\xi ) g_{n_{\beta_1}}(\bar{\beta}_1)g_{n_{\beta_2}}(\bar{\beta}_2)
g_{n_{\beta_3}}(\bar{\beta}_3) \chi_{K_1, n_{\eta_1}} (\eta_1 ) 
\chi_{K_2, n_{\eta_2}} (\eta_2 )\chi_{K_2, n_{\eta_2}} (\eta_2 ) 
\end{eqnarray}
where $K_k$ is the eigenvalue of the operator
$\frac{1}{i}\frac{\partial}{\partial\Phi_k}$ and $\tilde{K}$ stands for
$\mid K-K_1-K_2-K_3 \mid$.
Applying the transformations of the coordinate symmetries, as explained
in Ref. \cite{belly} and Appendix B,
the final form of the wave function,
given in Appendix B, is obtained.

The energy is

\begin{eqnarray}
E & = & \frac{\hbar^2}{2\Theta_{11}}[ I(I+1)-K^2] 
+ \hbar\omega_\eps (\mid K-K_1-K_2-K_3 \mid+2n_\eps
+1) \nonumber \\
& & + \hbar\omega_{\beta_1}(n_{\beta_1}+\frac{1}{2}) 
+ \hbar\omega_{\beta_2}(n_{\beta_2}+\frac{1}{2})     
+ \hbar\omega_{\beta_3}(n_{\beta_3}+\frac{1}{2}) \nonumber \\
& & + \hbar\omega_{\eta_1}(\frac{1}{2}\mid K_1\mid+ 2n_{\eta_1}+1) 
+ \hbar\omega_{\eta_2}(\frac{1}{2}\mid K_2\mid+2n_{\eta_2}+1) \nonumber \\  
& & + \hbar\omega_{\eta_2}(\frac{1}{2}\mid K_3\mid+2n_{\eta_3}+1) 
+ \hbar\omega_r (n_r+\frac{1}{2}) + \hbar\omega_\xi (n_\xi +\frac{1}{2})
~~~.
\label{energya}
\end{eqnarray}
The frequencies are

\begin{eqnarray}
& \hbar\omega_\eps  =  \hbar\sqrt{\frac{C_\eps}{\Theta_\eps}}
,~~~\hbar\omega_{\eta_k}  =  \hbar\sqrt{\frac{C_{\eta_k}}{B_k}}
,~~~\hbar\omega_{\beta_k}  =  \hbar\sqrt{\frac{C_{\beta_k}}{B_{\beta_k}}} &
\nonumber \\
& \hbar\omega_{r}  = 
\hbar\sqrt{\frac{C_r}{\mu_{(12)}}} ,~~~
\hbar\omega_{\xi}  =  
\hbar\sqrt{\frac{C_\xi}{\mu_{(12)3}}} ~~~. 
\end{eqnarray}

Before we apply the outlined procedure to the cases
$^{96}$Sr $+$ $^{10}$Be $+$ $^{146}$Be,
$^{112}$Ru $+$ $^{10}$Be $+$ $^{130}$Ru and $^{108}$Mo $+$ $^{10}$Be
$+$ $^{134}$Te, which are all non-symmetric systems,
the derivation of the potential parameters is outlined.
The ones related to the $\beta$ and $\gamma$ vibrations are obtained through
the spectrum of the individual clusters. For details see Refs.
\cite{belly,greiner}.

\vskip 1cm
\noindent

\section*{3. Derivation of relative potentials}

The aim of this section is to derive the expressions for the 
stiffness coefficients appearing in Eq.(\ref{potsin}). 
For that we need to calculate the interactions between the nuclei 
composing the trinuclear molecule. Since the interaction potential 
between the clusters should depend not only on their reciprocal distances 
but also on orientations we choose a double folding-model potential in 
which the nuclear densities $\rho$ are directly folded with effective 
nucleon-nucleon interactions :   
\begin{equation} 
V(\bd{R}) = \int d{\bd r}_{1} \int d{\bd r}_{2}~
\rho_{1} ({\bd r}_{1}) \rho_{2} ({\bd r}_{2}) v({\bd s}) 
\label{dfold}
\end{equation} 
For the nuclear potential we assume the M3Y effective interactions
\cite{bert77}. Like in previous papers \cite{double,prc98}, 
we consider only the isoscalar and isovector components for the 
central nucleon-nucleon force :
\begin{equation}
 v_{\rm M3Y}(\bd{s}\equiv \bd{R} + \bd{r}_{2} - \bd{r}_{1}) 
= v_{00} ( \bd{s}) +  \hat{J}_{00} \delta(\bd{s})+
(v_{01}(\bd{s})+\hat{J}_{01} \delta(\bd{s}))
\bd{\tau_{1}}\cdot\bd{\tau_{2}} 
\end{equation}

The isoscalar component of the M3Y force  is
 $$ v_{00} ( r ) = \left[ ~ 7999~ { e^{-4 r} \over 4 r }
 - 2134~ { e^{-2.5~ r} \over 2.5~ r } \right] ~~~ \mbox{MeV} $$
 and the isovector part has the form
 $$ v_{01} ( r ) = \left[ ~-4885.5~ { e^{-4 r} \over 4 r }
 + 1175.5~ { e^{-2.5~ r} \over 2.5~ r } \right] ~~~ \mbox{MeV} ~~\cdot $$
The strength of the isoscalar zero-range pseudo-exchange potential is taken 
according to the common prescription $\hat{J}_{00}$ = 
-276~MeV$\cdot$fm$^{3}$, whereas the isovector one is  $\hat{J}_{01}$ = 
217~MeV$\cdot$fm$^{3}$.

The ground state one-body nuclear densities of the fragments are taken as
Fermi distributions in the intrinsic frame
\beq
\rho ( \bd{r} ) = \rho_{0} \left[ 1 + \exp { 1 \over a } \left( r -
R_{0} \left( 1 + \sum_{\lam}
\beta_{\lam} Y_{\lam 0} (\theta,0 )
 \right) \right) \right]^{ -1}.
\label{fermi}
\eeq
The constant $\rho_{0}$ is fixed by normalizing the proton
and neutron density to the $Z$ proton and $N$ neutron
numbers, respectively. This condition ensures the volume conservation.
The radius $R_0$ and diffusivity parameters were taken from liquid
drop model calculations \cite{mol95} for the heavy fragments, whereas for 
the light cluster we consider the prescription $R_0=1.04\cdot A_3^{1/3}$ for 
the radius and $a=0.35$ for the diffusivity.
As static deformations, $\beta_\lam$, we considered quadrupole, octupole and
hexadecupole deformations.

However, due to the lack of an explicit density dependence in the 
M3Y effective interaction, this potential is characterized by a strong, 
unphysical attraction of a few thousands of MeV for total overlapping.
Therefore in the region of nuclear-density overlap we follow the suggestion
from \cite{seiw85,ueg93} and introduce a phenomenological repulsive 
potential which originates from the compression effects of the 
overlapping density
\beq
V_{\rm rep}(\bd{R}) = V_p\int d{\bd r}_{1} \int d{\bd r}_{2}~
\widetilde\rho_{1} ({\bd r}_{1}) \widetilde\rho_{2} ({\bd r}_{2}) 
\delta({\bd s}) \eeq
where the tildes on the densities signify a distribution of the same
shape as (\ref{fermi}) but possessing an almost sharp surface. 
The strength of the  compression term $V_p$ was determined from the nuclear 
equation of state  \cite{greiner} by requiring for total overlap of two
nuclei a double normal density of nuclear matter. For a given dinuclear
subsystem ($A_{\rm light}+A_{\rm heavy}$) we take the value of the nuclear 
compression modulus $K$ according to the receipt proposed in 
\cite{myswy98}, and compute $V_p$ from the equation giving the binding 
energy loss for total overlap \cite{bragin84}, i.e. $R=0$:
\beq
V_{\rm M3Y}(0) + V_{\rm rep}(0) \approx \frac{1}{9} K A_{\rm light}
\eeq

The double folded potential (\ref{dfold}) is computed by making
a general multipole expansion  for two final nuclei with distance $R$ between 
their center of masses and orientation in space given through the Euler angles 
$\Omega_{1}$  and  $\Omega_{2}$ 
\cite{ober83} 
\beq
 V (\bd{R} , \Omega_{1} , \Omega_{2} ) = \sum_{\lambda_{i},\mu_{i}}
\sqrt{\frac{4\pi}{2\lam_3+1}}
  V_{\lambda_{1} \lambda_{2} \lambda_{3}}^{\mu_{1} \mu_{2} \mu_{3}}
 ( R ) {D}_{\mu_{1} 0}^{\lambda_{1}} ( \Omega_{1} )
  {D}_{\mu_{2} 0}^{\lambda_{2}} ( \Omega_{2} )
  Y_{\lam_{3}\mu_{3}} ( \hat{R} ) 
\label{multipol}
\eeq
Details on the calculation of the potential multipolar components 
$ V_{\lambda_{1} \lambda_{2} \lambda_{3}}^{\mu_{1} \mu_{2} \mu_{3}}
( R )$ were given in \cite{double}.

In a previous paper \cite{letterdouble} we considered that when the 
nuclear molecule is bent the reciprocal distances between the heavy 
fragments and the light cluster are preserved, i.e. we allowed the 
trinuclear molecule to perform only vibrations which result in the 
increase of the angle between the two valence bonds. In this way we 
excluded possible bond stretching vibrations. This time we take into
account such a degree of freedom and  by $\delta r_{13(23)}$ we denote
the change in the distance between clusters 1(2) and 3. The geometry of
the system, presented in Fig.5 for a deformed light cluster, provides the
values of these quantities as a function of the deflection angles
$\ph_{1,2,3}$ :
\beqa
\delta r_{13} & = & - {1\over 2}\frac{R_1R_3}{(R_1+R_3)}(\ph_1+\ph_3)^2
\label{valbon13}
\nonumber\\
\delta r_{23} & = & - {1\over 2}\frac{R_2R_3}{(R_2+R_3)}(\ph_2-\ph_3)^2
\label{valbon23}
\eeqa
On the other hand, from the inspection of Fig.5, the interaction
(\ref{multipol}) between the deformed light cluster 3 and  the 
heavier deformed nucleus 1, reads : 
\beq
V (\bd{r}_{13} ) = \sum_{\lam_1,\lam_2,\lam_3}
\frac{4\pi}{\sqrt{(2\lam_1+1)(2\lam_2+1)}}
V_{\lam_1 \lam_2 \lam_3}^{\mu~-\mu~0} ( R_1+R_3+\delta r_{13} ) 
Y_{\lam_1 \mu}(\varphi_1-\eps,0) Y_{\lam_2-\mu}(\eps+\varphi_3,0) 
\label{pot13}
\eeq
Since we made earlier the approximation $\varphi_1\approx\eps$, the variation 
of the valence bond length (\ref{valbon13}) reads
\beq
\delta r_{13} =  - {1\over 8}\frac{R_1R_3}{R_1+R_3}
\left (\frac{R_1+R_2+2R_3}{R_1+R_3} \right)^2\eps^2
\eeq  
Expanding the potential (\ref{pot13}) with respect to the small angle
$\eps$ we have that
\beq
V(\bd{r}_{13}) = V(R_1+R_3) + {1\over 2}C_{\eps}^{13}\eps^2
\eeq
where 
\beq
C_{\eps}^{13} = -{1\over 4}\left ( \frac{R_0}{R_1+R_3}\right )^2
\sum_{\lam_i}\left( \frac{R_1R_3}{R_2+R_3}\frac{\partial
V_{\lam_1\lam_2\lam_3}^{0~0~0}(r_{13})}{\partial r_{13}}-
{1\over 2}\lam_2(\lam_2+1) V_{\lam_1\lam_2\lam_3}^{0~0~0}(r_{13})\right
)_{r_{13}=R_1+R_3}
\eeq
Using similar arguments we get the expression for the stiffness of the
butterfly mode, coming from the interaction between clusters 2 and 3
\beq
C_{\eps}^{23} = -{1\over 4}\left ( \frac{R_0}{R_2+R_3}\right )^2
\sum_{\lam_i}\left( \frac{R_2R_3}{R_2+R_3}\frac{\partial
V_{\lam_1\lam_2\lam_3}^{0~0~0}(r_{23})}{\partial r_{23}}-
{1\over 2}\lam_2(\lam_2+1) V_{\lam_1\lam_2\lam_3}^{0~0~0}(r_{23})\right
)_{r_{23}=R_2+R_3}
\eeq
The last contribution to the stiffness coefficient of the butterfly mode 
comes from the interaction of the heavier fragments 1 and 2.
Using again the geometry of Fig.5 we write in multipolar form the
interaction between these nuclei
\beq
V (\bd{r}) = \sum_{\lam_1,\lam_2,\lam_3}
\frac{4\pi}{\sqrt{(2\lam_1+1)(2\lam_2+1)}}
V_{\lam_1 \lam_2 \lam_3}^{\mu~-\mu~0} ( R_0 + \delta r ) 
Y_{\lam_1 \mu}(\varphi_1,0) Y_{\lam_2 -\mu}(\varphi_2,0) 
\label{pot12}
\eeq
According to eqs.(\ref{r13}-\ref{r12}) the shift in the interfragment 
distance reads
\beq
\delta r = -{1\over 2}R_0\frac{R_1+R_3}{R_2+R_3}\eps^2 + 
\delta r_{13} + \delta r_{23}
\eeq   
Consequently expanding in Taylor series this potential too, we obtain the
stiffness coefficient
{\small
\beqa
C_\eps & =  & {1\over 2}\sum_{\lam_1\lam_2\lam_3}
\left (\lam_1(\lam_1+1)\frac{R_1-R_2}{R_2+R_3}-
\lam_2(\lam_2+1)\frac{(R_1-R_2)(R_1+R_3)}{(R_2+R_3)^2}
\right.
\nonumber \\
& & \left.
~~~~~~~~~~~ - \lam_3(\lam_3+1)\frac{R_1+R_3}{R_2+R_3}\right )
V_{\lam_1\lam_2\lam_3}^{0~0~0}(R_0)
\nonumber\\              
&&-R_0\sum_{\lam_1\lam_2\lam_3}
\left ( \frac{R_1+R_3}{R_2+R_3}+\frac{R_0 R_3(R_1R_3+R_2R_3+2R_1R_2)}
             {4(R_1+R_3)(R_2+R_3)^3}\right )
\frac{\partial V_{\lam_1\lam_2\lam_3}^{0~0~0}(R_0)}{\partial r}
\eeqa       }

In order to obtain the stiffness coefficients of the bond streching vibrations 
$C_{13}$ and $C_{23}$, we expand the potentials $V(\bd{r}_{13})$ and 
$V(\bd{r}_{23})$ , up to second power of $\delta r_{13}$ and $\delta r_{23}$.
Such an expansion is possible in view of the relative minimum in the potential 
with respect to the inter-cluster distance. As an example we give in
Fig. 6
the  potential between the heavy cluster1(2)
and the light cluster 3.  
After some algebra we obtain for the dinuclear subensemble (13)
\beq
V(\bd{r}_{13}) = V(R_1+R_3) + A_{13}\delta r_{13} + 
{1\over 2}C_{13}\delta r_{13}^2
\eeq  
where
\beqa
A_{13} &=& \sum_{\lam_1,\lam_2,\lam_3}\left ( {\lam_2(\lam_2+1)\over 2}
\frac{R_1 R_3}{R_1+R_3}V_{\lam_1\lam_2\lam_3}^{0~0~0} (r_{13})
+ \frac{\partial V_{\lam_1\lam_2\lam_3}^{0~0~0}(r_{13})}{\partial r_{13}} 
\right )_{r_{13}=R_1+R_3}
\\
C_{13} &=& \sum_{\lam_1,\lam_2,\lam_3}
\left ( 
\frac{\lam_2(\lam_2+1)(3\lam_2^2+3\lam_2+1)}{24}
\left (\frac{R_1+R_3}{R_1R_3}\right )^2
V_{\lam_1\lam_2\lam_3}^{0~0~0}(r_{13})\right.\nonumber\\
& + &\left.\lam_2(\lam_2+1)\frac{R_1+R_3}{R_1R_3}
\frac{\partial V_{\lam_1\lam_2\lam_3}^{0~0~0} (r_{13})}{\partial r_{13}}
+\frac{\partial^2  V_{\lam_1\lam_2\lam_3}^{0~0~0}(r_{13})}{\partial 
r_{13}^2} \right )_{r_{13}=R_1+R_3}
\eeqa
Since the linear term in $\delta r_{13}$ has only the effect to shift the origin 
of the harmonic oscillator, the stiffness coefficient of the bond-stretching 
vibrations is specified by $C_{13}$. In the same manner we derive the coefficient 
$C_{23}$ and next using the relations from eq.(\ref{cr&cxi}) we derive the
stiffness coefficients of the $r$ and $\xi$-modes.

\section*{4. Applications to three-nuclear molecules}

The model is applied to the systems $^{96}$Sr $+$ $^{10}$Be $+$ $^{146}$Ba,
$^{90}$Y $+$ $^{10}$Be $+$ $^{142}$Cs and $^{108}$Mo $+$ $^{10}$Be
$+$ $^{134}$Te. For the computation of the numbers we use 
$mc^2$ = 938 MeV for the nuclear mass, $\hbar c$ = 197.33 MeV~fm$^2$, 
1.2 $A^{1/3}$ for the spherical equivalent radius of a heavy cluster and 
$1.3A^{1/3}$ for the light one. (Because we do not use here a diffuse 
surface as in the last section, for the spherical equivalent radii the usual 
formula $r_0A^{\frac{1}{3}}$ with $r_0=1.2$ for light and $1.3$ for heavy nuclei
is used.) In order to obtain $\frac{3\hbar^2}{J_{k}}$ we set it equal to the 
energy of the $2_1^+$ state of the individual clusters.

In case we consider a spherical nucleus, the $E(2_1^+)$ value is taken
as the vibrational energy. This has to be used with caution because for
a rotor the $2_1^+$ state is a rotational one. In a nuclear molecule
the rotation of individual clusters is constrained due to the link to
other clusters. Its rotation is converted into the butterfly motion.
This is not the case for a vibrational state. In this contribution we will
assume a deformed $^{10}$Be where the deformation is {\it not taken} from Ref.
\cite{raman} because the assumptions  used there are no longer valid for
light deformed nuclei. We rather use the $SU(3)$ model of the nucleus
and deduced from there a deformation of $0.175$ (see Ref. \cite{letter}).
It will be seen that the results do not sensitively depend on that.
Interpreting the $2_1^+$ state at $3.368$ MeV in $^{10}$Be as rotational,
it is absorbed into the butterfly motion. Instead the first vibrational
state is a $\gamma$-mode at $5.958$ MeV, indicating a very stiff system.
The fact that the $3.368$ MeV transition (minus the $6$ keV shift) is
seen in experiment \cite{hamilton,hamilton2} speaks in favour of a
vibrational $^{10}$Be nucleus. We carried out computations in frame of the 
Hartree-Fock method, with pairing correlations taken into
account and using Skyrme III forces. The result was that the deformation 
energy curve of $^{10}$Be has a spherical minimium and it is symmetric for 
small deformations. Nevertheless, we would like to see the effects of a 
possibly deformed light cluster. All this has to be taken into account when 
it comes to the interpretation of the theoretical results.

After having described the three systems mentioned above, shortly the
nuclear structure of the participants of the system
$^{90}$Y $+$ $^{10}$Be $+$ $^{142}$Cs will be discussed. $^{90}$Y and
$^{142}$Cs are odd-even nuclei with an odd number of protons. Their
treatment requires the inclusion of the spins of the extra protons.
Though, the extension can be done it would break the scope of this
contribution. In a subsequent paper it will be treated. The comments
we will make, however, are important for the interpretation of the
experimental results.

\subsection*{4.1 $^{96}$Sr $+$ $^{10}$Be $+$ $^{146}$Be}
The nuclear structure of this system was already discussed in Refs.
\cite{letter,letterdouble}.
The new contribution here is the inclusion of the
$\beta$ and $\gamma$ and the relative vibrational modes.
Although, the relative vibrational modes where discussed in Ref.
\cite{letter} numerical values where not given due to the missing
information on the stiffness parameters. These will be provided now.

The Sr and the Ba nuclei are prolately deformed and the
corresponding deformation values are given in Table 1.
As outlined in the last section and using the data from Ref. \cite{letter}
the parameters of the Hamiltonian are deduced and
listed in Table 1. The spectrum can then be determined from the Eq. (\ref{energya}).
The radii along the prolate symmetry axis are also listed in Table 1.
For the Be nucleus we showed in Ref. \cite{letter} that according to the $SU(3)$ 
model it can be taken as triaxial and the deformation is $0.175$. This 
consideration does exclude any mixing with other $SU(3)$ representations due to
the $SU(3)$ mixing terms like pairing and spin-orbit interaction.
There are many indications that the $^{10}$Be nucleus can be assumed to be 
spherical \cite{letter}. However, in order to see the influence of a deformed
light cluster we will assume a deformed $^{10}$Be nucleus.

Using the parameters of this system listed in Table 1, the spectrum
of the molecular states is given by (units are in MeV)

\begin{eqnarray}
E & = & 0.000862[ I(I+1)-K^2] 
+ 2.406(\mid K-K_1-K_2-K_3\mid +2n_\eps ) \nonumber \\
& & + 1.229 n_{\beta_1} 
+ 1.053 n_{\beta_2} + 6.179 n_{\beta_3}     \nonumber \\
& &
+ 1.507(\frac{1}{2}\mid K_1\mid+ 2n_{\eta_1})
+ 1.566(\frac{1}{2}\mid K_2\mid+ 2n_{\eta_2})
+ 5.958(\frac{1}{2}\mid K_3\mid+ 2n_{\eta_3})
\nonumber \\
& & + 3.61 n_r + 17.59 n_\xi ~~~,
\label{energy01}
\end{eqnarray}
where we have skipped the zero point energy contribution, i.e. $E$ gives
the difference in energy to the ground state.
All deformation vibrational states are lying above 1 MeV.
The same holds for the butterfly frequency. In conclusion,
below 1 MeV only the rotational states belonging to the ground state
band appear. The relative $\xi$ motion is at such a large energy that
it does not play any practical role. 
Interesting to note is that in the calculation of 
$\frac{2\Theta_{11}}{\hbar^2}$ the dominant
contribution comes from the
last
term of $\Theta_{11}$ as given
in Eq. (\ref{theta1}). The other terms contribute at most three percent.
Even the rotational contribution from the light Be cluster,
given by
the term before the last one
of $\Theta_{11}$, can be neglected
due to a small $\xi_0$.
This is similar to the two cluster case where the
corresponding term is dominating all others.

In Fig.7 the expected structure of the spectrum is plotted.
Only band heads contained in Eq. (\ref{energy01}) are
shown
and not
those belonging to other degrees of freedom, like the
rotational octupole band head
state
$1^-$ in $^{146}$Ba and the extra 0$^+$ state at 1.465 MeV in $^{96}$Sr.
As can be seen in Fig. 7, there are
no excited band head
states below 1 MeV suggesting a
stable behavior against the butterfly motion. Nearly all states below 1 MeV 
are rotational one belonging to the ground state band.
Note that in Ref. \cite{letter} The $\hbar\omega_\eps$ was
estimated assuming a spherical $^{10}$Be nucleus. 

The energy values of the first
$2_1^+$ and $4_1^+$ states are 5.2 keV and 17 keV respectively.
The deformation of $^{10}$Be results in an increase of the separation
of the heavy clusters which raises the moment of inertia. The expected
lowering in the energies of the rotational states is small compared
to the results of Ref. \cite{letter}, which are 6 keV and 20 keV for the
$2_1^+$ and $4_1^+$ states, indicating a small influence
of the supposed deformation of the $^{10}$Be nucleus.

\subsection*{4.2 $^{112}$Ru $+$ $^{10}$Be $+$ $^{130}$Sn}

This system is the most symmetric one we could get for which experimental
information about the structure of the individual clusters are available
and not just the ground state only. This is important for deducing
the deformation of the nuclei. This system has not been seen yet but
should exist.

The heavy fragments are again even-even nuclei. Using the tables of Ref.
\cite{raman} the
corrected
deformation of $^{112}$Ru is given by
$0.237$
corresponding to a large $\beta$. For $^{130}$Sn no information in
these tables are available. However, The Sn isotopes are known to
be an excellent example for the seniority scheme \cite{ring}. The proton
shell is closed and the neutron shell is open. Because the seniority scheme
is realized a zero deformation can be assumed. This is also confirmed by
the $(E(4_1^+)/E(2_1^+))$ ratio \cite{sakai} which is $1.63$ for $^{130}$Sn.
For $^{112}$Ru the ratio is $2.72$ indicating a rotational structure.

The parameters of the nuclei and the system are listed in Table 1, including
the radii of the clusters along the line of contacts. 
The excitation energy is given by (units are in MeV)

\begin{eqnarray}
E & = & 0.00094[ I(I+1)-K^2] 
+ 2.215(\mid K-K_1-K_2-K_3\mid+2n_\eps) \nonumber \\
& &
+ 0.524(\frac{1}{2}\mid K_1\mid+ 2n_{\eta_1})
+ 5.958(\frac{1}{2}\mid K_3\mid+ 2n_{\eta_3})
\nonumber \\
& & + 3.911 n_r + 19.035 n_\xi 
\nonumber \\
& & + 1.220 N_{\rm Sn}
~~~,
\label{energy02}
\end{eqnarray}
where the last term describes the five-dimensional harmonic oscillator
for the Sn nucleus. The spectrum is presented in Fig. 8,
with the same characteristics as in $^{96}$Sr $+$ $^{10}$Be $+$ $^{146}$Ba.
Not plotted are the band heads belonging to different degrees of freedom
than those described by the model (see discussion in the former
section). Also suppressed are the states of the five dimensional harmonic
oscillator of $^{130}$Sn with $\hbar\omega_{\beta}$ = 1.220 MeV.
There is a $\gamma$ vibrational state at approximately half a MeV though.
One should observe above this band head a rotational structure similar to
the ground state band with the difference that also a $3^+$ state exists
at about 5.5 keV and a $4^+$ state at 12.7 keV
above the $2^+$ band head state. The $\xi$ relative vibrational term can
also be neglected. In this system
the rotational part is dominated to
almost $100\%$ by the
last
term of $\Theta_{11}$ of Eq. (\ref{theta1}).
The influence of the assumed deformation of $^{10}$Be is again small.

\subsection*{4.3 $^{108}$Mo $+$ $^{10}$Be $+$ $^{134}$Te}
The heavy fragments are even-even nuclei. Using the tables of Ref.
\cite{raman} the deformation of $^{108}$Mo is given by $0.354$
which we corrected to $0.333$ using the additional deformation
dependent terms in the $B(E2;0_1^+ \rightarrow 2_1^+)$ as given
in Ref. \cite{greiner}. The experimental information of the spectrum
is taken from Ref. \cite{iso}. For $^{134}$Te no information in
these tables are available, however, the tendency observed coming from the
lighter isotopes indicates a small $\beta$. When we look at
the ratio $(E(4_1^+)/E(2_1^+))$ we obtain 2.92 and 1.23
for $^{108}$Mo and $^{134}$Te respectively.
This experimental observation supports a deformed nucleus for
$^{108}$Mo and possibly a spherical deformation for $^{134}$Te.

The $^{108}$Mo nucleus is particularly difficult to treat.
Within the geometrical model \cite{greiner}
the Potential-Energy-Surface (PES) of the neigbouring nucleus
$^{108}$Ru has a spherical
absolute minimum and a triaxial local minimum at large deformation
\cite{mo1}. The energy of the ground state, however, lies above
the saddle point and the ground state is a strong mixture between both
deformations. A large $\beta$ in average is also indicated by the
results of $\beta\beta$ decay for $^{108}$Mo
where with great success a model for
strongly deformed nuclei was applied \cite{mo2}.

Under the assumption that the Mo isotope is deformed while the Te
nucleus is a vibrator and using the parameters listed in Table 1,
the excitation energy is given by

\begin{eqnarray}
E & = & 0.00092[ I(I+1)-K^2] 
+ 2.127(\mid K-K_1-K_2-K_3\mid+2n_\eps) \nonumber \\
& &
+ 0.586(\frac{1}{2}\mid K_1\mid+ 2n_{\eta_1})
+ 5.958(\frac{1}{2}\mid K_3\mid+ 2n_{\eta_3})
\nonumber \\
& & + 3.54 n_r + 17.30 n_\xi 
\nonumber \\
& &
+ 1.280N_{Te}
~~~,
\label{energy03}
\end{eqnarray}
where the last term describes the five-dimensional harmonic oscillator
for the Te nucleus. The spectrum is presented in Fig. 9, again
with the same characteristics as in $^{96}$Sr $+$ $^{10}$Be $+$ $^{146}$Ba.
Not plotted are the band heads belonging to degrees of freedom
different from those of the model. Also excluded are the states of
the five dimensional harmonic oscillator of $^{134}$Te with
$\hbar\omega_{\rm Te}$ $=$ $1.280$ MeV.
There is a $\gamma$ vibrational state at approximately half a MeV though.
One should observe above this band head a rotational structure similar to
the ground state band with the difference that also a $3^+$ state exists
at about $5.5$keV and a $4^+$ state at $12.8$keV
above the $2^+$ band head state. The $\xi$ relative vibrational term can
also be neglected. In this system the rotational part is dominated to almost 
$100\%$ by the last term of $\Theta_{11}$ of Eq. (\ref{theta1}).
Again the interpretation with respect to the influence of the 
$^{10}$Be deformations similar as in the two former cases.

\subsection*{4.4 Structure of the participants in \\
$^{90}$Y $+$ $^{10}$Be $+$ $^{142}$Cs}

This is the third system
possibly
identified in the experiment of Ref.
\cite{hamilton2}.
The heavy clusters are odd-odd nuclei. The influence of the extra odd
protons should be included in our consideration. The way to do it
for individual nuclei is given in Ref. \cite{greiner}. It would,
however, exceed the scope of this contribution to add a corresponding
discussion. We refer to a later publication on this subject. Therefore,
we only discuss the nuclear structure of the participants. We consider
it important because the interpretation of the experimental results
rely heavily on the structure of the heavy clusters. They can be described by an 
even-even core and an odd proton around it \cite{greiner}.
The deformation can be deduced via the one of neighboring nuclei.
Take $^{90}_{38}$Sr$_{52}$, $^{90}_{40}$Zr$_{50}$ as neighboring
nuclei and also $^{88}_{38}$Sr$_{50}$ as the core. The deformation
of these nuclei are listed in Ref. \cite{raman}, except for
$^{90}_{38}$Sr$_{52}$. The deformation are $0.09$ and $0.12$ for
the last two nuclei respectively. However, one must take some care in
using these tables. The formula used to deduce the deformation
from the experimental $B(E2,0_1^+ \rightarrow 2_1^+)$ transition is theoretically 
biased, i.e. a strongly axial symmetric deformation is assumed. In this case the
deformation value is proportional to the square root of the transition.
For spherical and triaxial nuclei the formula is incorrect and in general
the deformations deduced are too high due to the fact that higher orders
\cite{greiner} in deformation are also neglected. Nevertheless, the tables
in Ref. \cite{raman} give a good idea about the trends.
Another possibility is to use the
tables of M. Sakai \cite{sakai} where the ratio $(E(4_1^+)/E(2_1^+))$
is investigated. For rotational nuclei the ratio is $3.33$ and for
vibrational nuclei it is $2.0$. Most nuclei lie in between these two values.
For the three nuclei mentioned above this ratio is $1.99$, $1.41$ and
$3.15$ respectively. Except for the last value, which is near to a rotator,
the data indicate a spherical deformation, in agreement with the data listed in 
Ref. \cite{raman}.

For $^{142}_{55}$Cs$_{87}$ as neighboring nuclei we took $^{142}_{54}$Xe$_{88}$, 
$^{142}_{56}$Ba$_{86}$ and for the core $^{140}_{54}$Xe$_{86}$.  The deformation 
values listed in Ref. \cite{raman} are $0.157$ and $0.114$ for the last two 
systems. No information is listed for the first nucleus.
Using the tables of Sakai \cite{sakai}, except for the first nucleus were we used 
the ISOTOPE EXPLORER \cite{iso}, the ratio
of $(E(4_1^+)/E(2_1^+))$ are respectively $2.41$, $2.32$ and $2.22$.
They hint to a spherical nucleus while the $\beta$ values
are in between.
In Ref. \cite{hamilton2} a deformed nucleus was assumed, which are not
confirmed by data.

\section*{5. Conclusions}
In this contribution we proposed a model for three cluster molecules,
whose existence are suggested by experiments 
in Refs. \cite{hamilton,hamilton2}. First results
of their theoretical description
are published
in Refs. \cite{letter,letterdouble}.
The modes of this molecule are the butterfly mode,
rotations of the whole system and $\beta$ and $\gamma$ vibrations of
each cluster. The belly-dancer mode is a particular rotation around the
molecular $z$-axis. In general, all clusters can be deformed.

In order to render the model soluble, assumptions had to be introduced.
First, the three clusters are supposed to form linear chain molecule with the
lightest cluster in the middle. Second,
the inclination of the symmetry axis of
the clusters and the distance of the lightest cluster to the molecular
$z$ axis were assumed to be small. For later applications we
restrict the consideration to the case where
the first cluster touches the third one and this the second nucleus.
Relative vibrations were also added but taken to be in the direction of
the $z$ axis. Even with these assumptions, for large asymmetric
systems there is a coupling between the butterfly motion (denoted by
the angle $\eps$) and the rotation around the molecular $y$ axis, which
is perpendicular to the $(x,z)$ plane in which the center of masses of
the three clusters are located. For "nearly" symmetric clusters and
when the ratio $\frac{m_3}{m_1+m_2}$ is very small, this coupling can 
also be neglected. When there is a stretching motion along the molecular
$x$ axis, coupling terms appear which have to be diagonalized in
the basis constructed in this contribution.

One basic feature of the model is that the butterfly motion is alway possible
and the contributions come from the motion of the center of masses of
the three clusters, contrary to the two cluster case where the deformation
plays the essential role for $\hbar\omega_\eps$. The contribution
to $\frac{C_{\eps}}{\hbar^2}$ is more than half from the
center of mass terms in Eq. (\ref{theta1}) and less than half coming from
the deformation dependent terms. In the last terms the $^{10}$Be
contribution is negligible.

The model was applied to the systems $^{96}$Sr $+$ $^{10}$Be $+$ $^{146}$Ba,
$^{122}$Sn $+$ $^{10}$Be $+$ $^{120}$Ru,
and $^{108}$Mo $+$ $^{10}$Be $+$ $^{134}$Te.
All these splittings have the characteristic that there are few vibrational states 
below 1 MeV. The last two systems exhibit a low lying $\gamma$ vibrational
band head at about 0.5 MeV.
The rotational states of the ground state band are strongly squeezed.
The first excited $2^+$ state is at approximately 5-6 keV and the $4_1^+$
state at approximately 17-19 keV. The butterfly motion lies at energies
larger than 3 MeV indicating a particular strong stiffness against
this mode, contrary to the two cluster molecule.
The structure, obtained for these systems,
have to be considered as a rough approximation, due to the assumptions made
and the neglection of couplings between most of the degrees of freedoms.

The structure obtained for individual nuclei of the last system
and for $^{90}$Y $+$ $^{10}$Be $+$ $^{142}$Cs, where
only a discussion
on the nuclear structure of the individual clusters is given, are
in contradiction to Ref. \cite{hamilton2}. There, for example the
$^{142}$Cs nucleus is assumed to be deformed while experimental data prefer
a spherical nucleus in its ground state. Also the nucleus $^{108}$Mo
is taken as spherical in Ref. \cite{hamilton} while the data suggest
a deformed nucleus. Because for the three cluster molecule the main
contributions come from their center of masses, changes in
the deformation of the nuclei will not affect the main results.
However, in Ref. \cite{hamilton2} the deformation plays an important
role because it explains the
possible
systematics observed of the shift in the
energy of the $2_1^+$ state in $^{10}$Be. Based on the structural
investigation we made, the explanation of the shift is probably not as easy.
We suggest explicit microscopic investigations of the influence of a
heavy cluster to the smaller one, maybe implying a change in structure
of also the heavy fragments from deformed to spherical and vice versa.
Also a revision of the experimental analysis is recommended.

\section*{Acknowledgments}
P.O.H. acknowledge funding from DAAD, DGAPA and CONACyT and
\c S.M. the financial support from DAAD and DFG (Germany).

\section*{APPENDIX A: \\ The scalar product $(\dot\xi^{lab}\cdot\dot\xi^{lab})$ 
in terms of the velocities $\omega_i^\prime$ and $\dot\eps$}

As spherical components we use the expression
\begin{eqnarray}
\xi_{\pm 1} & = & \pm \frac{1}{\sqrt{2}}(\xi_x \pm \xi_y) \nonumber \\
\xi_0 & =& \xi_z
\end{eqnarray}
where the definition of $\xi_{\pm 1}$ differs from the usual one
\cite{edmonds} for convenience. In the molecular frame the vector
$\bd{\xi}$ is defined to lie in the molecular plane given by the
molecular $z$ axis, parallel to the vector $\bd{r}$. Therefore,
the $\xi_y$ component vanishes and the relation of the cartesian to the
spherical components is such that the $\xi_{\pm 1}$ components are
in their absolute value identical and are given by $\pm\frac{1}{\sqrt{2}}\xi_x$.

Neglecting tems of the order $O(\eps^3)$ in Eq.(9) we obtain 
\beqa
\xi_x & = & (R_1+R_3)\eps\nonumber \\
\xi_z & = & \xi_0 +{1\over 2}
\frac{R_1+R_3}{R_2+R_3}\frac{A_1R_2-A_2R_1}{A_1+A_2}\eps^2
\eeqa
where $\xi_0$ is the value of the $\xi$-coordinate in the linear chain 
configuration (see eq.(10)).
The time derivative of $\xi^{lab}_m$ is determined using the procedure
outlined in Ref. \cite{greiner}. After some algebra, which implies also the 
calculation of Wigner $D$-functions time derivatives we arrive at 

\begin{eqnarray}
(\dot\xi^{lab}\cdot\dot\xi^{lab})\equiv \sum_m(-1)^m \xi^{lab}_m\xi^{lab}_{-m} 
&\approx&  
(R_1+R_3)^2\dot\eps^2 +\xi_0^2(\omega_1^{\prime 2} + \omega_2^{\prime 2}) +
(R_1+R_3)^2\eps^2\omega_3^{\prime ~2}
\nonumber \\ 
&+& 2 \xi_0 (R_1+R_3)(\dot\eps \omega_2^{\prime} -
\eps\omega_1^{\prime}\omega_2^{\prime}) ~~~.
\end{eqnarray}


\section*{APPENDIX B: \\ Coordinate symmetries of the tri-nuclear \\
molecule with prolate nuclei}

We follow the procedure outlined in the Refs. \cite{belly,greiner}.
The transformations of the molecular frame consists of the operators
$\hat{R}_{1,m}$ and $\hat{R}_{2,m}^2$, the $m$ refers to the molecular
frame, and their action is given by

\begin{eqnarray}
\hat{R}_{1,m} (x,y,z) & = &  (x,-y,-z) \nonumber \\
\hat{R}_{2,m} (x,y,z) & = &  (-x,-y,z) ~~~. \nonumber \\
\end{eqnarray}
These operators act on the Euler angles $\vartheta_k$ ($k=1,2,3$), the
other coordinates $\chi_i$, $\phi_i$, $\Phi_i$ ($i=1,2,3$) and
$\bar{r}$ and $\xi$. The result is given in Table 2. The angles $\chi_i$, $\phi_i$
and $\Phi_i$ correspond to the Euler angles describing the rotation of
the nucleus from the molecular frame to the principal axis of cluster no.
$i$.

Because the three clusters are supposed to lie in a plane the angles $\chi_i$
are put to zero. Also a small butterfly angle is assumed and all angles
$\varphi_i$ are proportional to it. A coordinate symmetry transformation
consists of those actions where the angels $\chi_i$ and $\varphi_i$ are changed
at most by a sign. Inspecting Table 2, the only combinations allowed are
$\hat{R}_{1,m}\hat{R}_{1,p_i}$ and $\hat{R}_{2 p_i}$ whose action
is given in Table 3.

These operators have to be applied to the solution

\begin{eqnarray}
& &\phi (\vartheta_1, \vartheta_2, \vartheta_3, \eps , \bar{r}, \bar{\beta}_1,
\bar{\beta}_2,\bar{\beta}_3,\eta_1,\eta_2,\eta_3) =  
e^{i(K_1\Phi_1+K_2\Phi_2+K_3\Phi_3)} 
D^{I^*}_{MK} (\vartheta)\chi_{\tilde{K}, n\eps} (\eps) g_{n_r} (\bar{r})
\nonumber \\
& & \times g_{n_\xi}(\xi ) g_{n_{\beta_1}}(\bar{\beta}_1)g_{n_{\beta_2}}(\bar{\beta}_2)
g_{n_{\beta_3}}(\bar{\beta}_3) \chi_{K_1, n_{\eta_1}} (\eta_1 ) 
\chi_{K_2, n_{\eta_2}} (\eta_2 )\chi_{K_2, n_{\eta_2}} (\eta_2 ) 
\end{eqnarray}
of the Schr\"odinger equation. The action of $\hat{R}_{2,p_i}$ ($i=1,2,3$)
on this state changes $\Phi_i$ to $\Phi_i\pm\frac{\pi}{2}$ and
$\eta_i$ to $-\eta_i$, i.e. it acts only on $e^{iK_i\Phi_i}$ and
$\chi_{K_i,n_{\eta_i}}(\eta_i)$. The result is a phase $(-1)^{K_i}$
implying only even values of $K_i$.

The action of the operator $\hat{R}_{1,m}$ $\hat{R}_{1,p_i}$ is more
involved. The result is

\begin{eqnarray}
& &\phi (\vartheta_1, \vartheta_2 \vartheta_3, \eps , \bar{r}, \bar{\beta}_1,
\bar{\beta}_2,\bar{\beta}_3,\eta_1,\eta_2,\eta_3) =  \nonumber \\
& &{\cal N} \left\{ D^{I~*}_{M~K} (\vartheta) 
f(\Phi_1,\Phi_2,\Phi_3)  + (-1)^{I-K}D^{I~*}_{M~-K} (\vartheta) 
f(-\Phi_1,-\Phi_2,-\Phi_3) \right\}  \nonumber \\
& &\times\chi_{K, n\eps} (\eps)g_{n_r} (\bar{r}) g_{n_\xi}(\xi ) 
g_{n_{\beta_1}}(\bar{\beta}_1 )g_{n_{\beta_2}}(\bar{\beta}_2)g_{n_{\beta_3}}(\bar{\beta}_3) 
\chi_{K_1, n_{\eta_1}} (\eta_1 ) 
\chi_{K_2, n_{\eta_2}} (\eta_2 )\chi_{K_2, n_{\eta_2}} (\eta_2 ) 
\nonumber\\
&&
\end{eqnarray}
where ${\cal N}$ is a normalization factor and
\begin{eqnarray}
f(\Phi_1,\Phi_2,\Phi_3) & = & e^{i(K_1\Phi_1+K_2\Phi_2+K_3\Phi_3)}
+ e^{-i(K_1\Phi_1+K_2\Phi_2-K_3\Phi_3)} \nonumber \\
& & e^{i(K_1\Phi_1-K_2\Phi_2-K_3\Phi_3)}+
e^{-i(K_1\Phi_1-K_2\Phi_2+K_3\Phi_3)} ~~~.
\end{eqnarray}

The quantum numbers acquire the possible values

\begin{eqnarray}
K_i & = & 0, 2, 4, ... \nonumber \\
K & = & 0, 1, 2, 3, 4, ... \nonumber \\
L & = & K, K+1, K+2, ...  \nonumber \\
n_r,~ n_\xi, ~ n_{\eps},~ n_{\beta_i},~ n_{\eta_i} & = & 0,1,2,3,4, ... ~~~.
\end{eqnarray}

\newpage
\begin{center}
FIGURE CAPTIONS:
\end{center}
\vskip 1cm
\noindent
Figure 1: Illustration of the main variables of a three cluster molecule.
The light cluster is plotted as spherical. In general it will be deformed
as the heavy nuclei. \\

\vskip 0.5cm
\noindent
Figure 2: In the upper half the mode discussed in this paper is plotted.
The lower half presents a more complex case which corresponds to the
anti butterfly mode in the two nuclear molecule for
the limit of vanishing mass of the light cluster.

\vskip 0.5cm
\noindent
Figure 3: The potential of the light cluster 3 in the field of the two
heavier fragments along the molecular $z$ axis, for three fixed
tip distances : $d$= 2 (solid lines), 3 (dashed line), 4 (dotted
line) FM. The trinuclear molecule comprises $^{96}$Sr $+$
$^{10}$Be $+$ $^{146}$Ba.

\vskip 0.5cm
\noindent
Figure 4: Main variables for the discussion of the motion of the
clusters with respect to each other. Only the centers of mass of the
nuclei are plotted. \\

\vskip 0.5cm
\noindent
Figure 5: In case the third cluster is deformed, the alignment of
a heavy nucleus with the light one is not perfect. The relations
of the angles are illustrated in this figure. \\

\vskip 0.5cm
\noindent
Figure 6: The potential between the heavy fragment $^{146}$Ba  and 
$^{10}$Be (solid line) and between $^{96}$Sr and $^{10}$Be (dashed line).

\vskip 0.5cm
\noindent
Figure 7: Spectrum of the system $^{96}$Sr $+$ $^{10}$Be $+$ $^{146}$Ba.
For detailed explanations, see the text. Only band heads are shown.
On top of each band head there is a rotational band with the characteristics
explained in the text. The butterfly mode is the $1^+$ state to the right
and the first relative vibration is given by the $1^-$ state. The $2^+$
states belong to the $\gamma$ vibrational states with either $K_1$ or
$K_2$ equal to $2$. The $0^+$ band heads consist of the ground state,
$\beta$ and $\gamma$ vibrational (with $n_{\eta_i}=1$)
band heads of the heavy clusters.

\vskip 0.5cm
\noindent
Figure 8: Spectrum of the system $^{112}$Ru $+$ $^{10}$Be $+$ $^{130}$Sn.
On top of each band head there is a rotational band with the characteristics
explained in the text. The butterfly mode is the $1^+$ state to the right
and the first relative vibration is given by the $1^-$ state. The $2^+$
state belongs to the $\gamma$ vibrational state of the first cluster
with
$K_1=2$. The $0^+$ band heads consist of the ground state and the
$\gamma$ vibrational (with $n_{\eta_i}=1,...,4$) band heads of
$^{112}$Ru. No $\beta$ vibrational states could be identified in the three
clusters at energies below 5 MeV.

\vskip 0.5cm
\noindent
Figure 9: Spectrum of the system $^{108}$Mo $+$ $^{10}$Be $+$ $^{134}$Te.
On top of each band head there is a rotational band with the characteristics
explained in the text. The butterfly mode is the $1^+$ state to the right
and the first relative vibration is given by the $1^-$ state. The $2^+$
state belongs to the $\gamma$ vibrational state of the first cluster with
$K_1=2$. The $0^+$ band heads consist of the ground state and the
$\gamma$ vibrational (with $n_{\eta_i}=1,...,3$) band heads of
$^{108}$Mo. No $\beta$ vibrational states could be identified in the three
clusters at energies below 5 MeV.

\newpage
\begin{table}
\begin{tabular}{|c|c|c|c|}
\hline
system & $^{96}Sr + ^{10}Be + ^{146}Ba$ & $^{112}Ru +^{10}Be + ^{130}Sn$ & $^{108}Mo + ^{10}Be + ^{134}Te$ \\
\hline 
$\beta_{01}$ & 0.338 & 0.237   & 0.333 \\
$\beta_{02}$ & 0.199 &  0. &  0.  \\
$\beta_{03}$ & 0.175 & 0.175 & 0.175 \\
$\frac{\hbar^2}{2\Theta_{11}}$ & 8.62 $10^{-4}$ & 9.10 $10^{-4}$ & 9.20 $10^{-4}$   \\
$\hbar\omega_{r}$ & 4.270    &  3.911   & 3.850   \\
$\hbar\omega_{\xi}$ &  19.417   &   19.035  & 18.511     \\
$\hbar\omega_{\eps}$ & 2.427 & 2.880 & 2.681 \\
$\hbar\omega_{\beta_1}$ & 1.229    &  -   & -   \\
$\hbar\omega_{\beta_2}$ & 1.053    & 1.220   & 1.280   \\
$\hbar\omega_{\beta_3}$ & 6.179    & 6.179    & 6.179    \\
$\hbar\omega_{\eta_1}$ & 1.507    & 0.524    &  0.586    \\
$\hbar\omega_{\eta_2}$ & 1.566   &  -   &  -   \\
$\hbar\omega_{\eta_3}$ & 5.958   & 5.958   &  5.958   \\
$R_1$ &  6.66 & 6.88  & 6.92 \\
$R_2$ & 7.11 & 6.08  & 6.14 \\
$R_3$ & 3.11 & 3.11 & 3.11 \\
\hline
\end{tabular}
\caption{Parameters of the three systems discussed in this paper. In case
the {\it deformation} $\beta_k$, for one particular nucleus $k$, is
{\it zero} the
$\hbar\omega_{\beta_k}$ has to be interpreted as the energy $\hbar\omega_k$
of the five dimensional harmonic oscillator. For the case of an oscillator the
corresponding $\hbar\omega_{\eta_k}$ is put to zero because it is not
relevant. The deformation parameters have no units. The one of the
$\hbar\omega_k$ are in MeV. The units of the radii are
is in fm, where we used for the spherical equivalent radius
the formula $r_0A^{1/3}$ with $r_0=1.2$ for a heavy and $r_0=1.3$ for the
light cluster. An "-" indicates that either no information is available,
very insecure or $\beta_{0k}$ is zero.
}
\end{table}

\newpage
\begin{table}
\begin{tabular}{|c|c|c|c|c|}
\hline
$variable$ & $\hat{R}_{1,m}$ & $\hat{R}^2_{2,m}$ & $\hat{R}_{1,p_i}$ &
$\hat{R}_{2,p_i}$ \\
\hline 
$\vartheta_1$ & $\vartheta_1+\pi$ & $\vartheta_1$     & $\vartheta_1$ & $\vartheta_1$ \\
$\vartheta_2$ & $\pi-\vartheta_2$ & $\vartheta_2$     & $\vartheta_2$ & $\vartheta_2$ \\
$\vartheta_3$ & $-\vartheta_3$    & $\vartheta_3+\pi$ & $\vartheta_3$ & $\vartheta_3$ \\
$\chi_i$   & $-\chi_i$      & $\chi_i+\pi$   & $\chi_i$   & $\chi_i$   \\
$\varphi_i$   & $\pi -\varphi_i$   & $\varphi_i$       & $\varphi_i+\pi$& $\varphi_i$   \\
$\Phi_i$   & $\Phi_i+\pi$   & $\Phi_i$       & $\pi -\Phi_i$ & $\Phi_i+\frac{\pi}{2}$ \\
$\xi_i$    & $\xi_i$        & $\xi_i$        & $\xi_i$   & $\xi_i$  \\
$\eta_i$   & $\eta_i$       & $\eta_i$       & $\eta_i$  & $-\eta_i$   \\
$\bar{r}$  & $\bar{r}$      & $\bar{r}$      & $\bar{r}$ & $\bar{r}$  \\
$\xi$      & $\xi$          & $\xi$          & $\xi$     &$\xi$        \\
\hline
\end{tabular}
\caption{The action of the basic coordinate symmetry operators on the
collective variables}
\end{table}

\newpage
\begin{table}
\begin{tabular}{|c|c|c|}
\hline
$variable$ & $\hat{R}_{1,m}\hat{R}_{1,p_i}$ & $\hat{R}_{2,p_i}$ \\
\hline 
$\vartheta_1$ & $\vartheta_1+\pi$ & $\vartheta_1$     \\
$\vartheta_2$ & $\pi-\vartheta_2$ & $\vartheta_2$     \\
$\vartheta_3$ & $-\vartheta_3$    & $\vartheta_3$     \\
$\chi_i$   & $-\chi_i$      & $\chi_i$        \\
$\varphi_i$   & $-\varphi_i$      & $\varphi_i$        \\
$\Phi_i$   & $-\Phi_i$      & $\Phi_i+\frac{\pi}{2}$  \\
$\xi_i$    & $\xi_i$        & $\xi_i$         \\
$\eta_i$   & $\eta_i$       & $\eta_i$         \\
$\bar{r}$  & $\bar{r}$      & $\bar{r}$       \\
$\xi$      & $\xi$          & $\xi$           \\
\hline
\end{tabular}
\caption{The action of the allowed combinations of symmetry operators
which satisfy the condition that after their application $\chi_i$ is
still zero and $\varphi_i$ is maintained near zero.}
\end{table}

\newpage

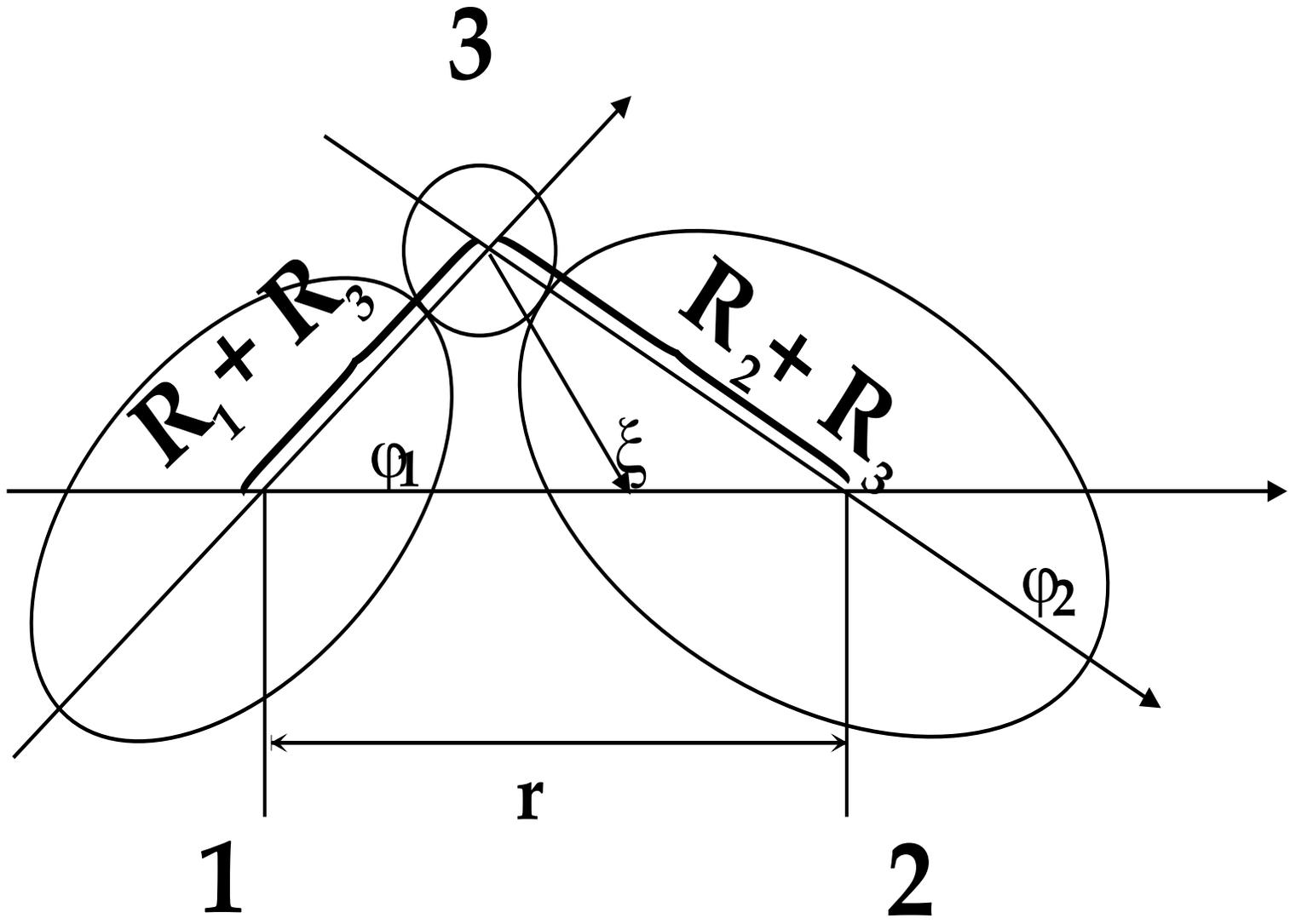
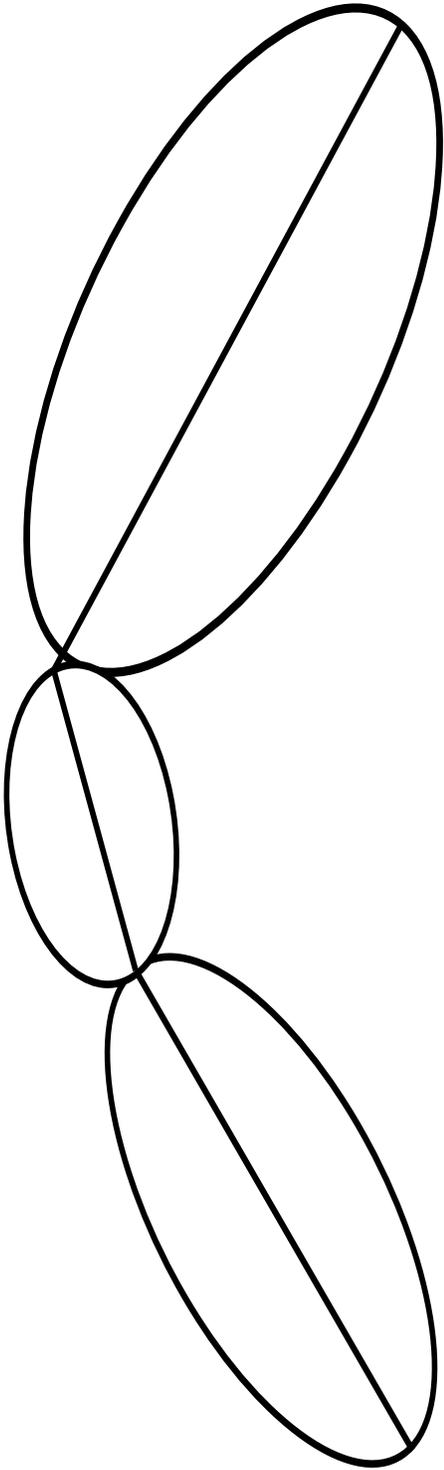
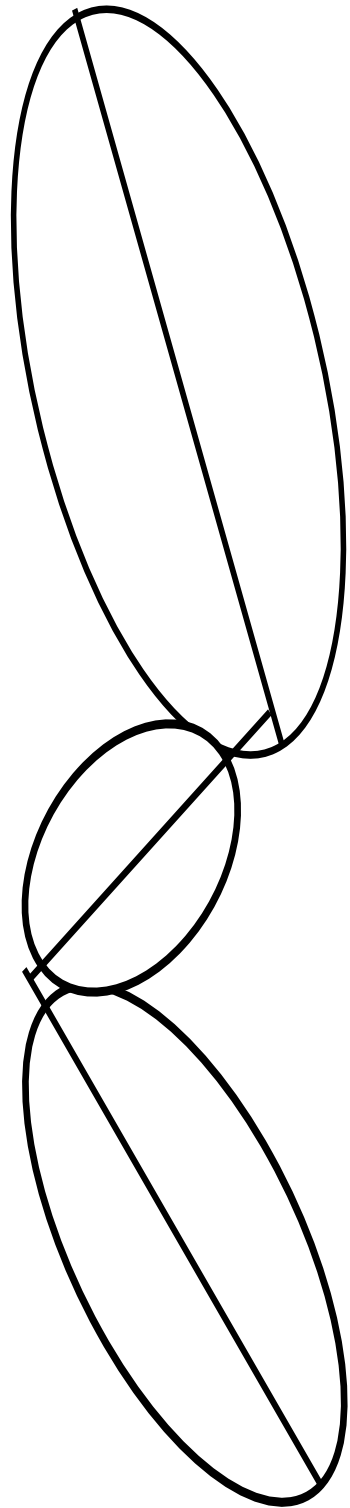
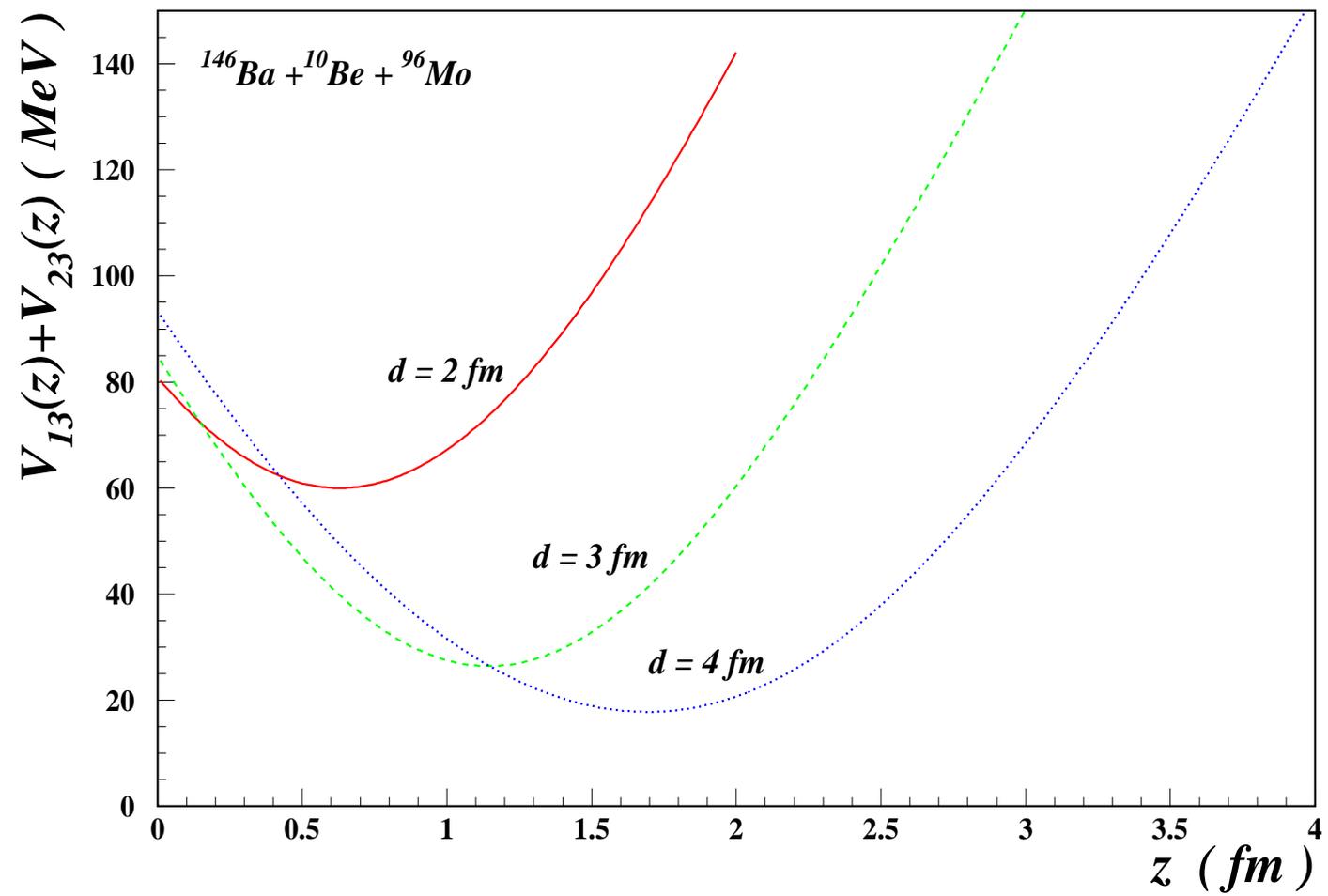
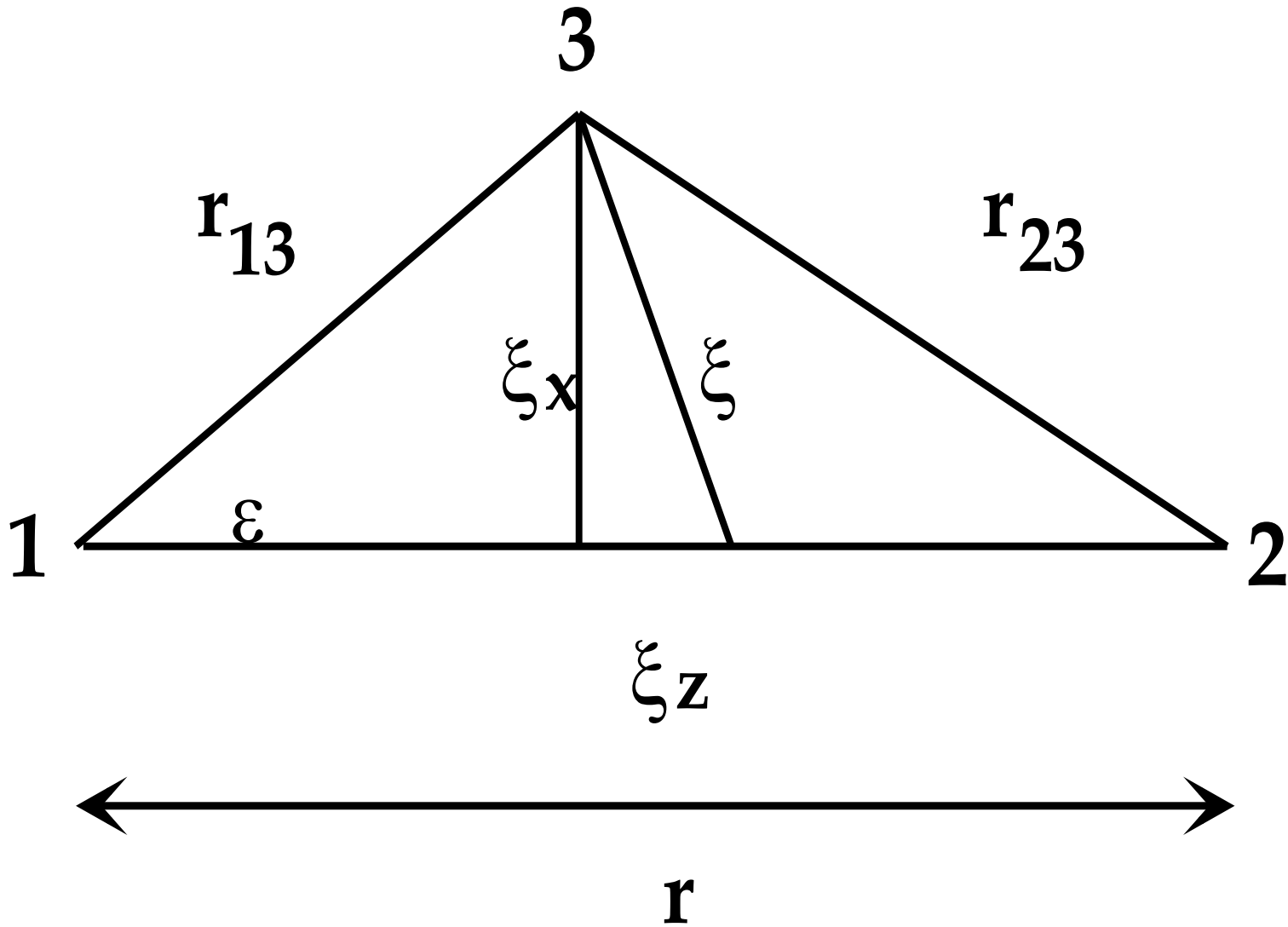
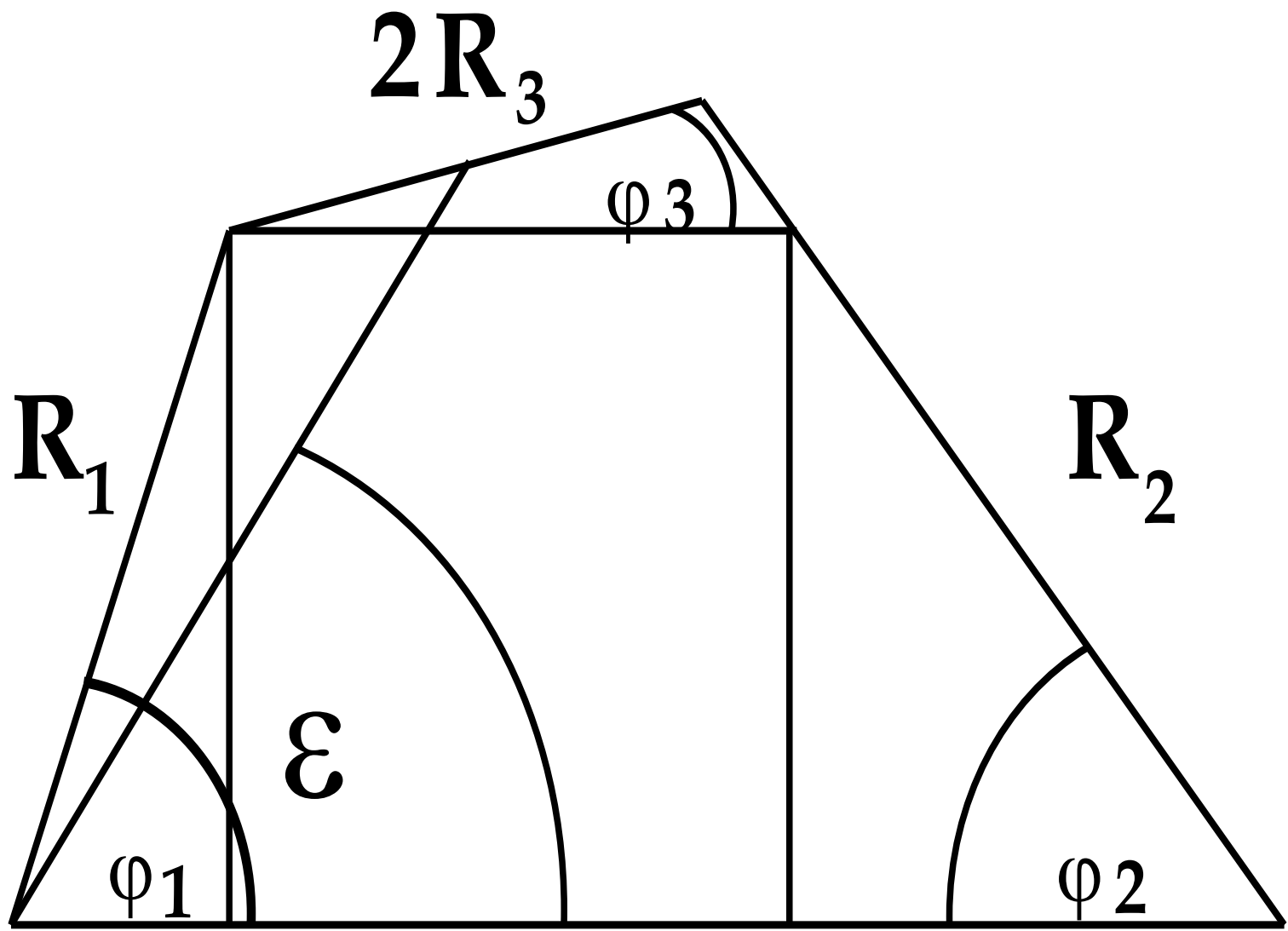
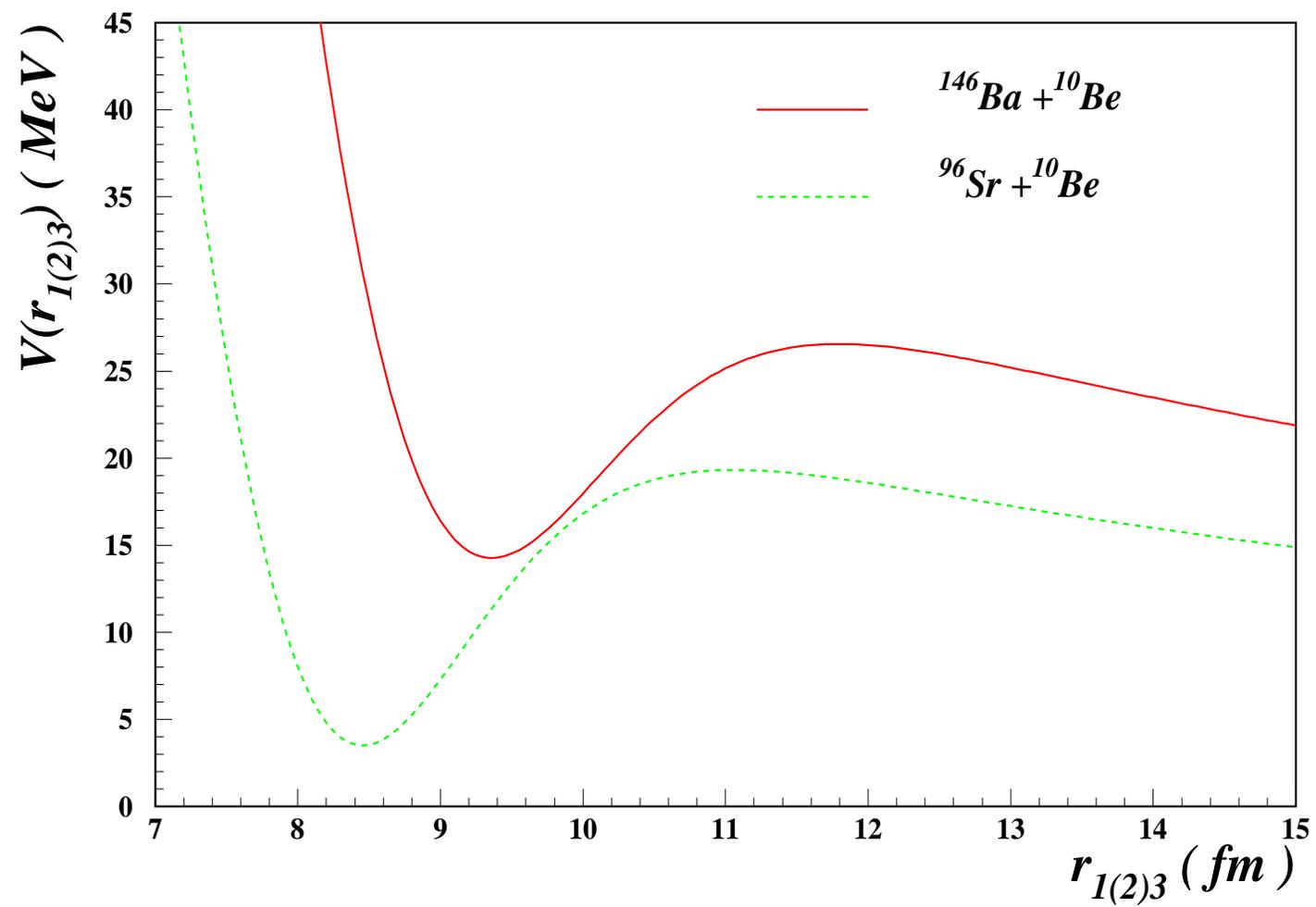
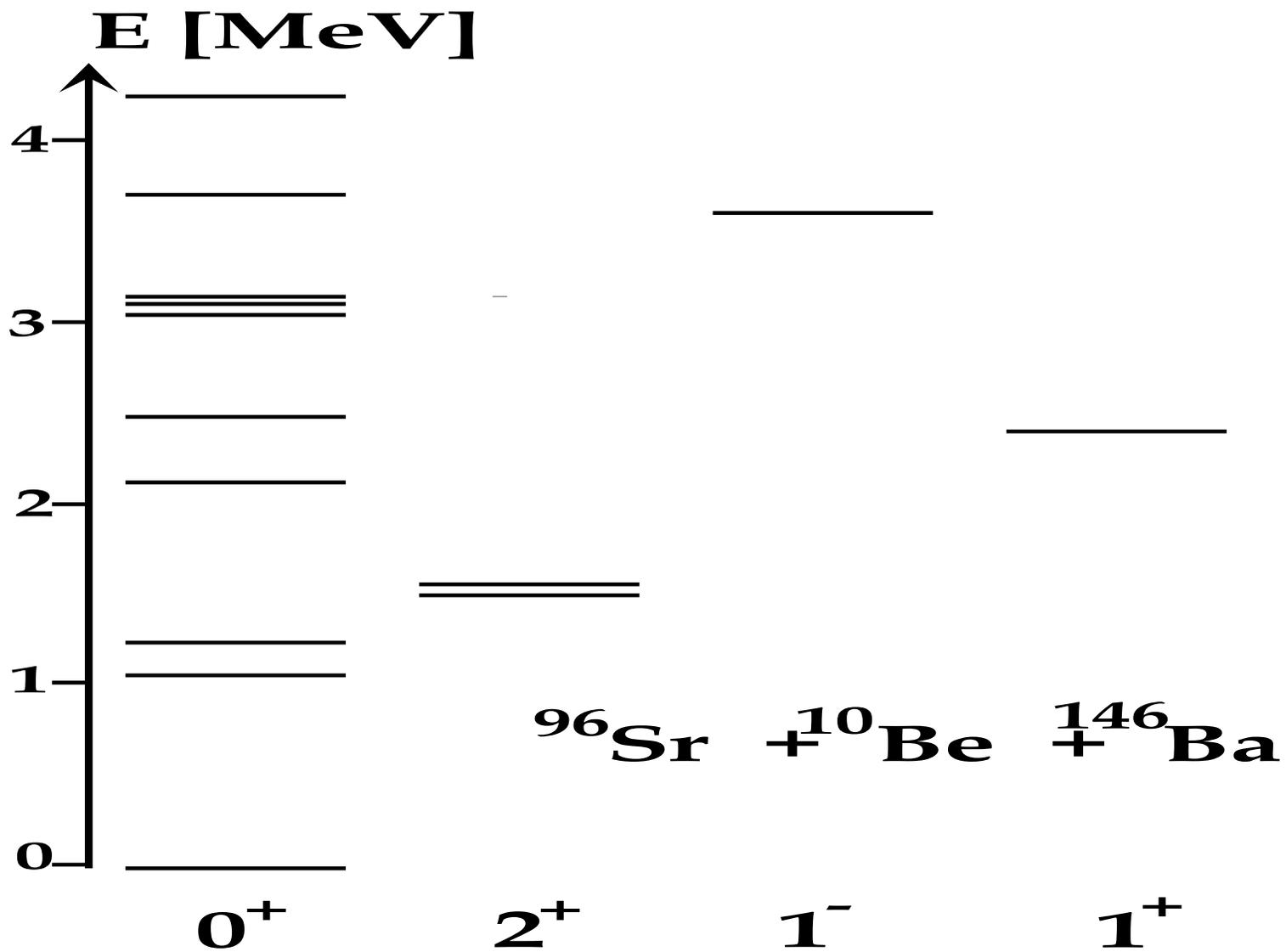
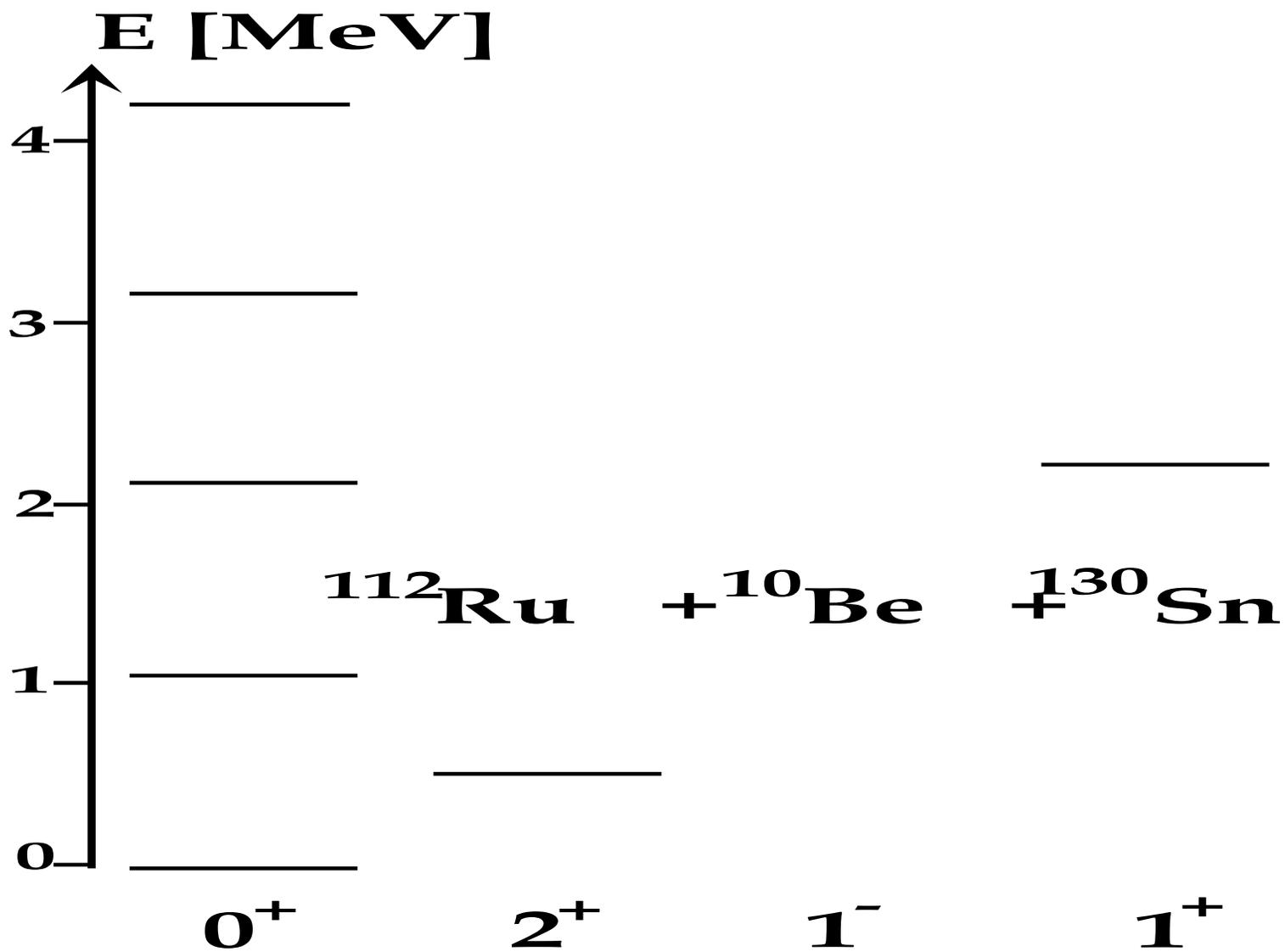
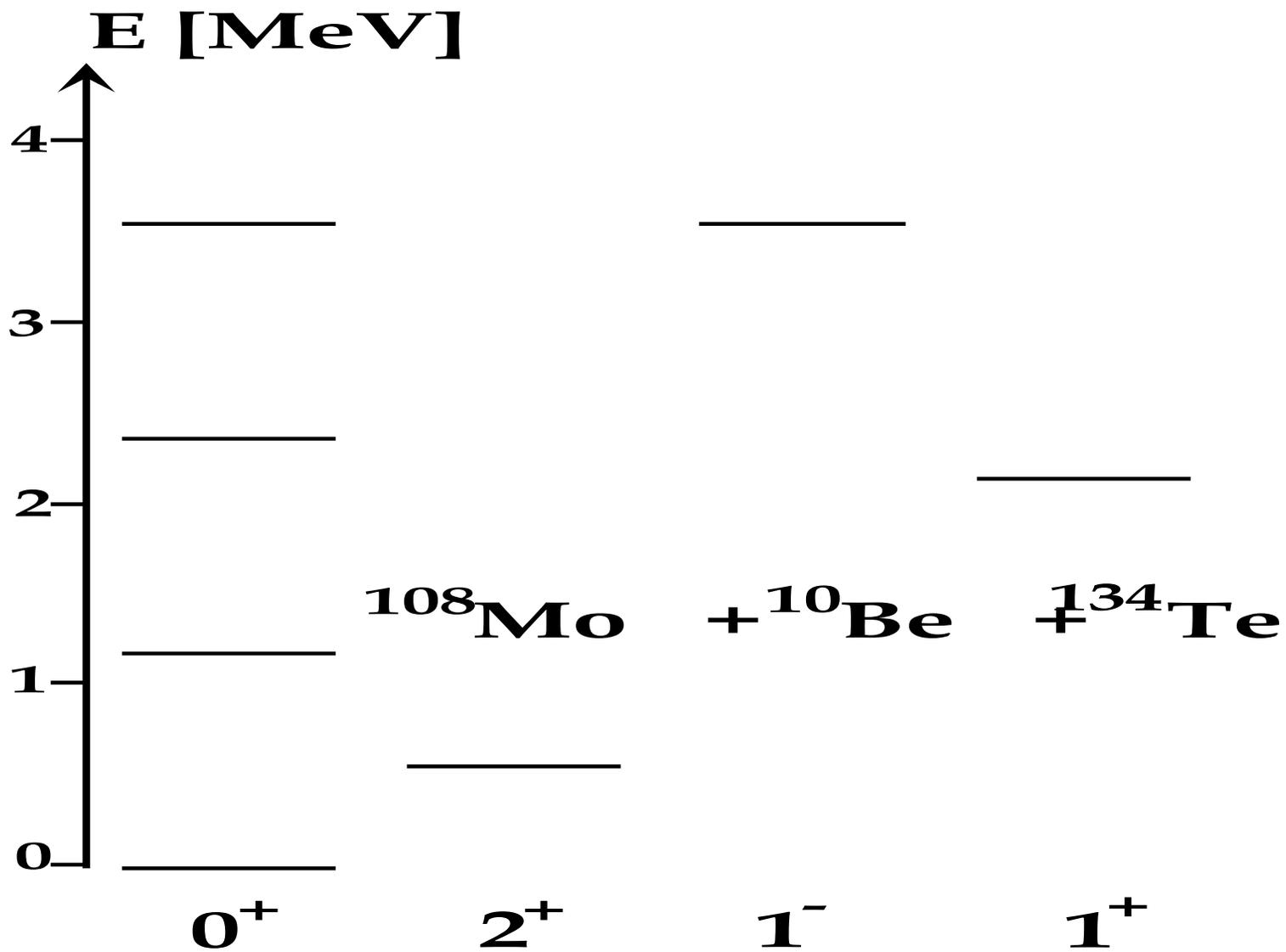
\end{document}